\documentclass[aps,pre,twocolumn]{revtex4}
\usepackage{hyperref}
\usepackage{color}
\usepackage{graphics,graphicx,epsfig,wrapfig}
\usepackage{epsf,epstopdf}
\usepackage{amssymb,amsfonts,amsmath}
\usepackage{enumitem}

\usepackage{ifthen}

\newcommand\mydots{\hbox to 0.8em{.\hss.\hss.}}
\newcommand{\beqn}{\begin{eqnarray}}
\newcommand{\eeqn}{\end{eqnarray}}
\newcommand{\beq}{\begin{equation}}
\newcommand{\eeq}{\end{equation}}

\newcommand{\degree}{{}^{\rm o}}

\newcommand{\ff}{1}

\newcommand{\mytitle}{Persisting fetal clonotypes influence the structure and overlap of adult human T cell receptor repertoires}

\newcommand{\myauthors}{Mikhail V. Pogorelyy$^{1}$, Yuval Elhanati$^{2}$, Quentin Marcou$^{2}$, \\
Anastasia L. Sycheva$^{1}$, Ekaterina A. Komech$^{1}$, Vadim I. Nazarov$^{1}$,\\
Olga V. Britanova$^{1,3,4}$,
Dmitriy M. Chudakov$^{1,3,4}$, Ilgar Z. Mamedov$^{1}$,\\
Yuri B. Lebedev*$^{1}$,
Thierry Mora*$^{5}$ and Aleksandra M. Walczak*$^{2}$}

\begin{document}
\title{\mytitle}
\author{\myauthors}
\affiliation{~\\
\normalsize{$^{1}$ Shemyakin-Ovchinnikov Institute of Bioorganic
  Chemistry,}\\
\normalsize{Moscow, Russian Federation}\\
\normalsize{$^{2}$ Laboratoire de physique th\'eorique,}\\
\normalsize{CNRS, UPMC and \'Ecole normale sup\'erieure, Paris, France}\\
\normalsize{$^{3}$Pirogov Russian National Research Medical University, Moscow, Russian Federation}\\
\normalsize{$^{4}$Masaryk University, Central European Institute of Technology, Brno, Czech Republic}\\
\normalsize{$^{5}$ Laboratoire de physique statistique,}\\
\normalsize{CNRS, UPMC and \'Ecole normale sup\'erieure, Paris,
  France}\\
\normalsize{\rm *Equal contribution}\\
}

\date{\today}

\begin{abstract} 
The diversity of T-cell receptors recognizing foreign pathogens is generated through a highly stochastic recombination process, making the independent production of the same sequence rare. Yet unrelated individuals do share receptors, which together constitute a ``public'' repertoire of abundant clonotypes. The TCR repertoire is initially formed prenatally, when the enzyme inserting random nucleotides is downregulated, producing a limited diversity subset. 
By statistically analyzing deep sequencing T-cell repertoire data from twins, unrelated individuals of various ages, and cord blood,
we show that T-cell clones generated before birth persist and maintain high abundances in adult organisms for decades, slowly decaying with age. 
Our results suggest that large, low-diversity public clones are created during pregnancy, and survive over long periods, providing the basis of the public repertoire.

\end{abstract}

\maketitle

The adaptive immune system relies on the diversity of T-cell receptor (TCR) repertoires to protect us from many possible pathogenic threats. This diversity is produced by a V(D)J recombination machinery that assembles the repertoire {\it de novo} in each individual, adding a large degree of randomness to combinations of genomic templates. The diversity is encoded not only in the set of specific receptors expressed in a given individual, but also in their relative abundances (the size of T-cell clones expressing each of them) which can differ by orders of magnitude. These differences are partially attributed to antigenic stimulation (infection, vaccination), implying that clones increase their sizes in response to common or recurring infections. Despite this great diversity,
different individuals---regardless of their degree of relatedness---do express a subset of the exact same receptors, called the {\it public} repertoire \cite{Venturi2006}. This overlap is often interpreted as the convergence of individual repertoire evolutions in response to common antigenic challenges \cite{Madi2014a}.
Accordingly, some public TCRs are known to respond for common pathogens such as the cytomegalovirus or Epstein-Barr virus \cite{Miles2011}.
However, this interpretation is challenged by the observation that these two properties---large differences in clone sizes and public repertoires---are also observed in naive repertoires, for which antigenic stimulation is not expected to be important \cite{Neller2015,Moon2007203}.

An alternative explanation for public clones, which does not invoke convergent repertoire evolution, is
that both abundant and public receptors are more likely to be produced by rearrangement, and just occur by coincidence \cite{Venturi:2011co, Venturi2006}. This idea is backed by some compelling evidence. 
First, the extent of sharing of clonotypes between pairs of individuals can be accurately predicted in both naive and memory pools from statistical models of sequence generation \cite{Elhanati2014}. 
Second, the likelihood that a clonotype sequence is shared by individuals has been reported to correlate with its abundance \cite{Zvyagin2014,Venturi:2011co}. However the origin of this correlation remains elusive. 
In addition, public clonotypes often have few or no randomly inserted N nucleotides, which limits their diversity.
Terminal deoxynucleotidyl transferase (TdT), the enzyme responsible for random nucleotide insertions, is not active in invariant T-cell subsets \cite{Venturi2013} and in some fetal T-cell clones, and these subsets could contribute to the emergence of the public repertoire. Another confounding factor is the ageing of repertoires, and the concomitant loss of diversity, which is expected to affect the structure of clonal abundances as well as the repertoire's sharing properties. 
How do all these effects shape the structure and diversity of TCR repertoires, and control their functional capabilities?
{Here we propose and test the hypothesis that a sizeable fraction of public clonotypes are created before birth. These clonotypes have low diversity because of reduced TdT activity, making them more likely to be shared among unrelated invididuals. Their large abundances, due to reduced homeostatic pressures in the early stages of repertoire development, allows them to survive over long periods.}

\begin{figure}
\noindent\includegraphics[width=\ff\linewidth]{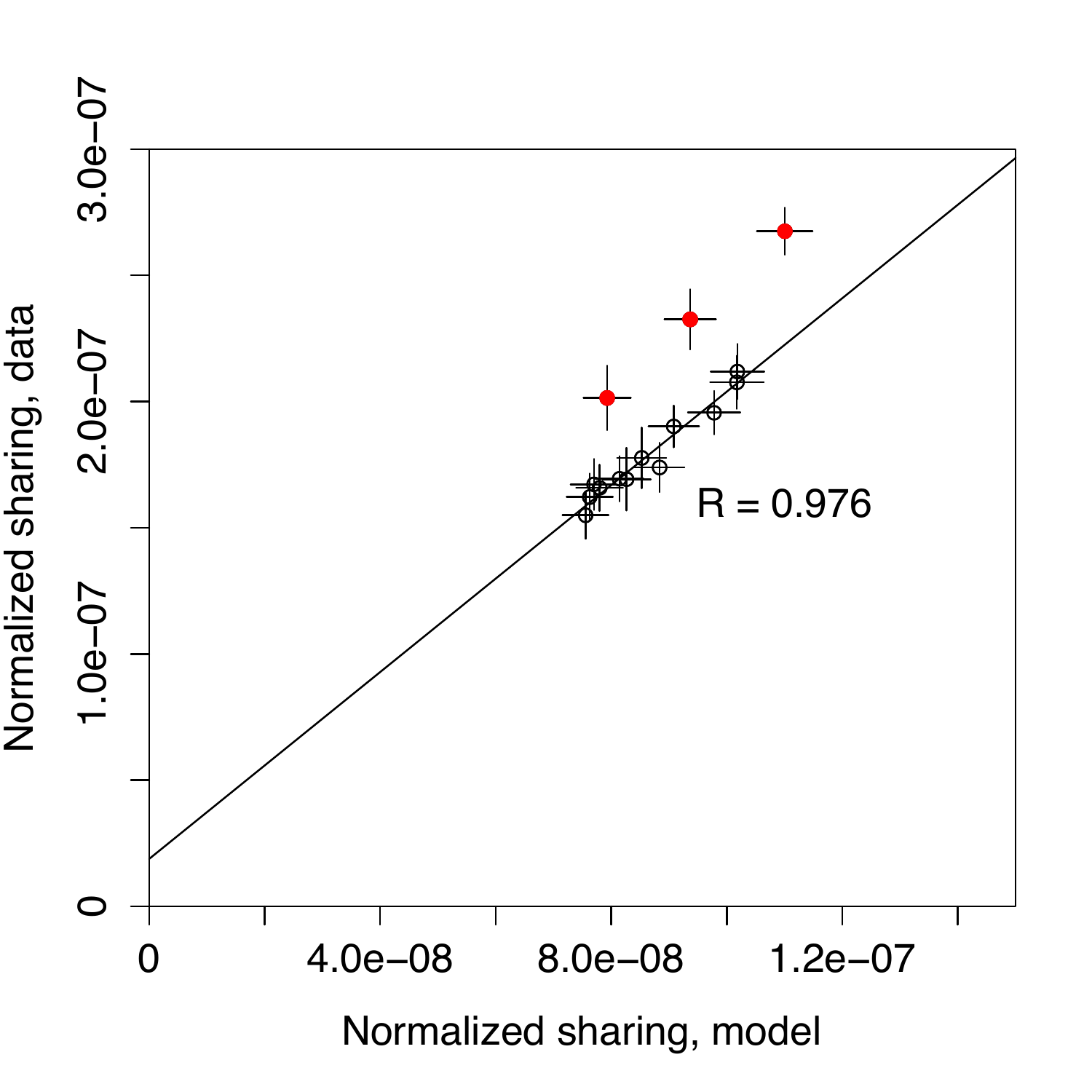}
\caption{{\bf TCR out-of-frame repertoire sharing in monozygous twins is higher than in unrelated individuals, or than predicted by stochastic models of recombination.} The number of shared out-of-frame alpha TCR clonotypes between all 15 pairs among 6 donors consisting of 3 twin pairs (ordinate) is compared to the model prediction (abscissa). To be able to compare pairs of datasets of different sizes, the sharing number was normalized by the product of the cloneset sizes. The three outstanding red circles represent the twin pairs, while the black circles refer to pairs of unrelated individuals. The model prediction is based on a generative stochastic model of VJ recombination \cite{Elhanati2015b}, inferred separately for each donor to account for differences between individuals. It agrees well with the data from unrelated individuals {up to a common multiplicative factor}, but systematically underestimates sharing in twins. Error bars show one standard deviation.}
\label{fig1}
\end{figure}

We first examined in detail the question of clonotype sharing between individuals. To avoid confounding effects due to convergent selection, we first focused on out-of-frame receptor sequences resulting from unsuccessful recombination events, which give us direct insight into the raw V(D)J recombination process \cite{Robins2010,Murugan2012}, free of clonal selection effects. The number of shared clonotypes between two clonesets is approximately proportional to the product of the cloneset sizes \cite{Murugan2012,Zvyagin2014,Shugay2013}. We call the ratio of the two the normalized sharing number.
Under the assumption that sharing occurs by pure chance, this number can be predicted
using data-driven generative probabilistic models of V(D)J recombination accounting for the frequencies of the assembled V, D, and J gene segments and the probabilities of insertions and deletions between them \cite{Murugan2012,Elhanati2014,Elhanati2015b}. We can estimate sharing either of the entire nucleotide chain (alpha or beta), or of a restricted portion of it called the Complementarity Determining Region 3 (CDR3), which concentrates most of the chain's diversity and determines antigen specificity.
Genetically identical individuals may be expected to have more similar recombination statistics \citep{Zvyagin2014,Glanville2011,Wang2014}, and therefore share more sequences.
To assess these genetic effects, we looked at the sharing of TCR alpha and beta-chain receptor repertoires between three pairs of monozygous twins. We synthesized cDNA libraries of TCR alpha and beta chains from the donors' peripheral blood mononuclear cells and sequenced them on the Illumina HiSeq platform (see Fig.~S1 and SI Text).
For each pair of individuals, the normalized number of shared 
out-of-frame alpha
sequences was 
compared to the prediction from the recombination model, as shown in Fig.~\ref{fig1} (see also Figs.~S2 for similar results on sharing of CDR3 sequences).
{Sharing in unrelated individuals  (black circles) was well predicted by the model (Pearson's $R=0.976$), up to a constant multiplicative factor of 2.07, probably due to differences in effective cloneset sizes.} While twins did share more sequences than unrelated individuals (red circles), this excess could not be explained by their recombination process being more similar. The model prediction was obtained by generating sequences from models inferred using each individual's cloneset as input \cite{Elhanati2015b}, mirroring their specific recombination statistics (see SI Materials and methods). The normalized sharing number departed significantly from the model prediction only in twins, 
calling for another explanation than coincidence in that case.
{The same result was obtained for beta out-of-frame CDR3 sequences (Fig.~S3), although less markedly because of a lower signal-to-noise ratio due to smaller numbers of shared sequences. Most of beta out-of-frame sequences shared among the highest-sharing twin pair associated with CD8 CD45RO+ (memory) phenotype in both individuals. This observation is surprising, because the non-functionality of these sequences excludes convergent selection as an explanation for it (see SI for details).}

We then examined the sharing of in-frame CDR3 sequences.
Most of in-frame sequences are functional, and have passed thymic and peripheral selection. Since these selection steps involve HLA types and are therefore expected to be similar in related individuals, we wondered whether the functional repertoires of twins also displayed excess sharing. Remarkably, we found some excess sharing in the in-frame beta
repertoire
(Fig.~S4), but none
in the in-frame alpha 
reperoire (Fig.~S5).
However, the failure to observe excess sharing in this last case can be explained by the much higher expected number of shared sequences in the alpha in-frame repertoire (due to both in-frame sequences being more numerous than out-of-frame ones, and to the lower diversity of alpha chains compared to beta chains) which could mask this excess in twins (see SI Text). 

\begin{figure}
\noindent\includegraphics[width=\ff\linewidth]{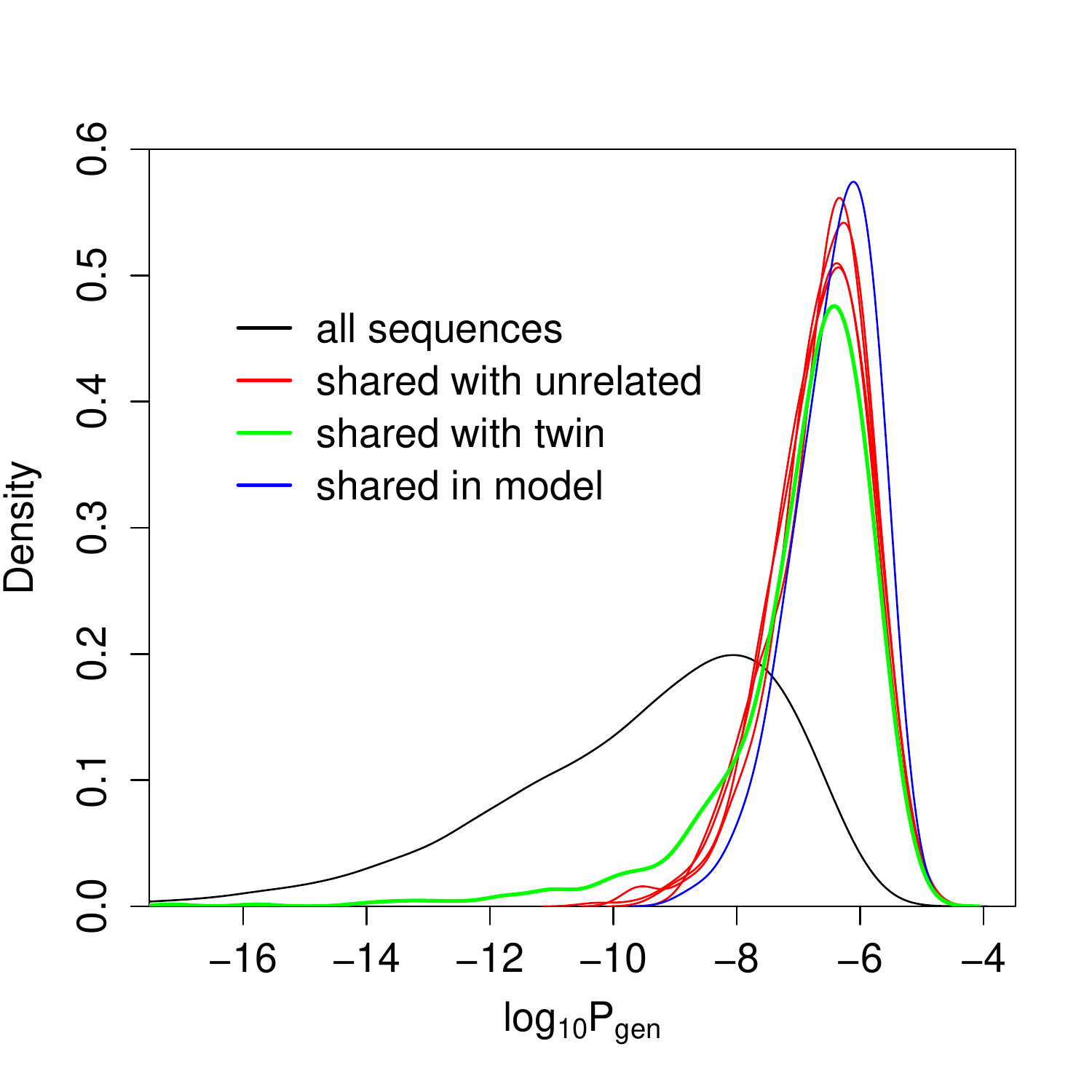}
\caption{{\bf TCR sequences shared between twins are statistically different from sequences shared between unrelated individuals.} Distribution of $\log_{10}P_{\rm gen}$, with $P_{\rm gen}$ the probability that a sequence is generated by the VJ recombination process, for shared out-of-frame TCR alpha clonotypes between one individual and the other five.
While the distribution of shared sequences between unrelated individuals (red curves) is well explained by coincidental convergent recombination as predicted by our stochastic model (blue), sequences shared between two twins (green) have an excess of low probability sequences. For comparison the distribution of $P_{\rm gen}$ in regular (not necessarily shared) sequences is shown in black.
}
\label{fig2}
\end{figure}

To investigate the origin of excess sharing between twins, we looked at the statistical properties of shared alpha out-of-frame sequences from Fig.~1. Shared clonotypes between non-twins, which happen by coincidence, should have a higher probability $P_{\rm gen}$ to have been produced by V(D)J rearrangement compared to non-shared clonotypes. Indeed, the distribution of $P_{\rm gen}$ among shared sequences can be calculated from the probabilistic model of generation (Fig.~\ref{fig2}, blue curve), and the prediction agrees very well with the data between non-twins (red curves). By contrast, shared sequences between twins deviate from the prediction (green curve), especially in the tail of low-probability sequences, but are consistent with a mixture of $18\pm 3\%$ of regular sequences (black curve), and the rest of coincidentally shared sequences (blue curve). These numbers agree well with the excess sharing in twins, which amounts to $17\%\pm 3\%$ of non-coincidentally shared sequences, as estimated from Fig.~1.
Sequences shared between twins also have higher numbers of insertions and are therefore longer than those shared between unrelated individuals or according to the model {(Fig.~S6)} -- a trend that is even more pronounced in memory cells (Fig.~S7).
Note these observations about recombination probabilities and number of insertions are related: sequences with many insertions each have a low generation probability because of the multiplicity of inserted nucleotides.

Taken together, these observations support the existence of another source of shared sequences than coincidence in twins.
Since the sharing of cord blood between twins is the only natural instance when the immune systems of two individuals share cells,
we propose that the increased sharing of private TCRs between identical twins dates back to the sharing of cord blood cells, and that these shared clones persist into late age. 

\begin{figure*}
\noindent\includegraphics[width=.38\linewidth]{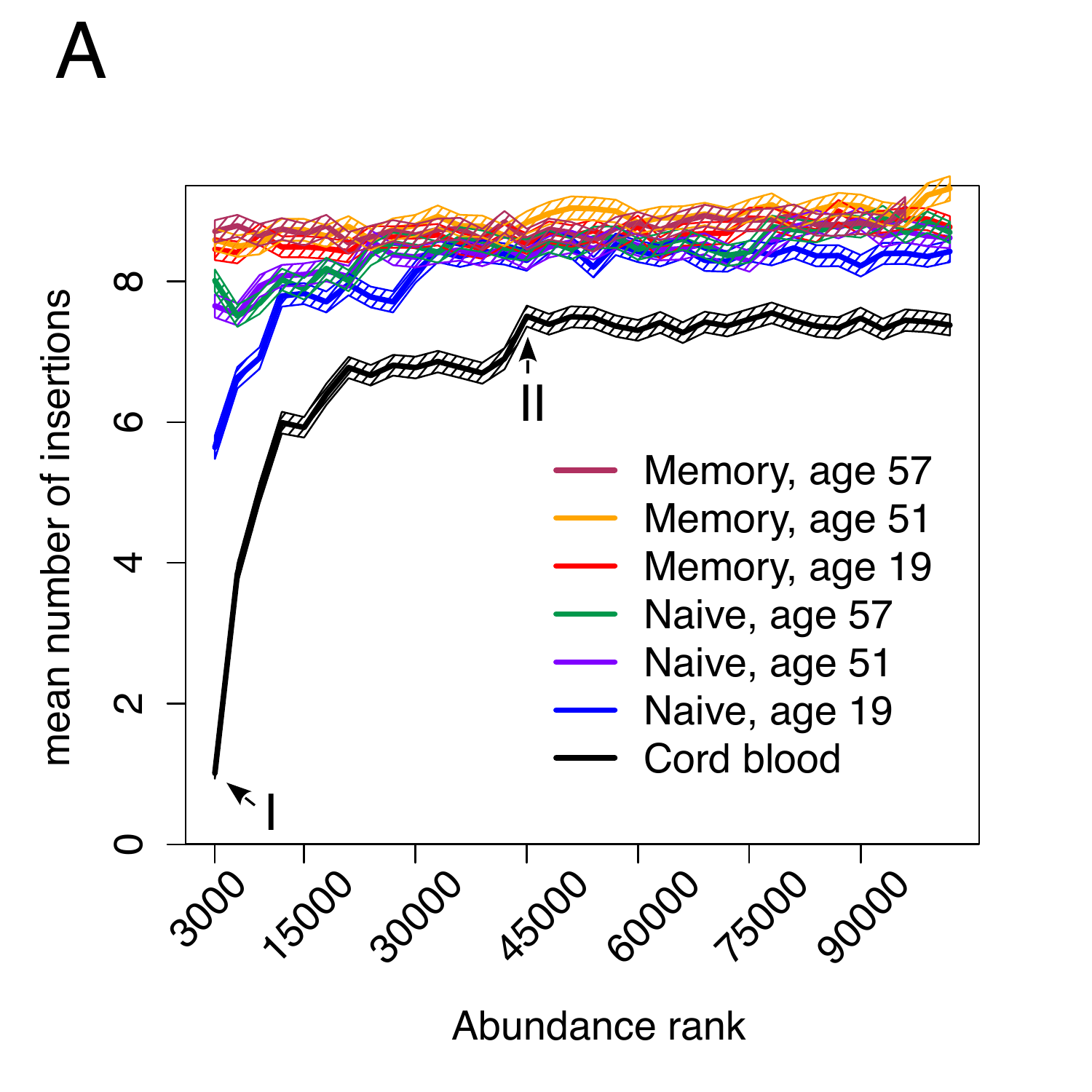}
\includegraphics[width=.219\linewidth]{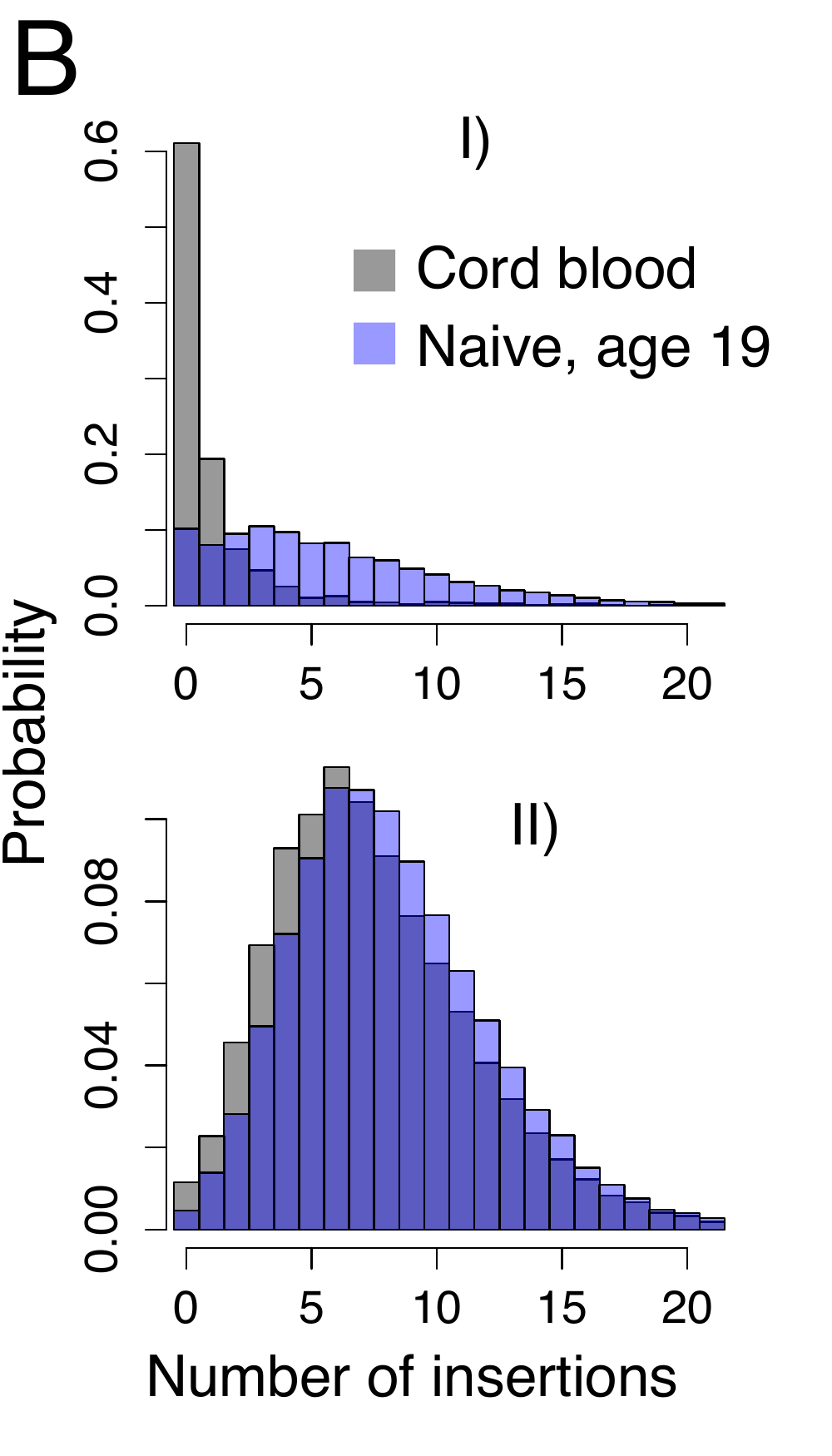}
\includegraphics[width=.385\linewidth]{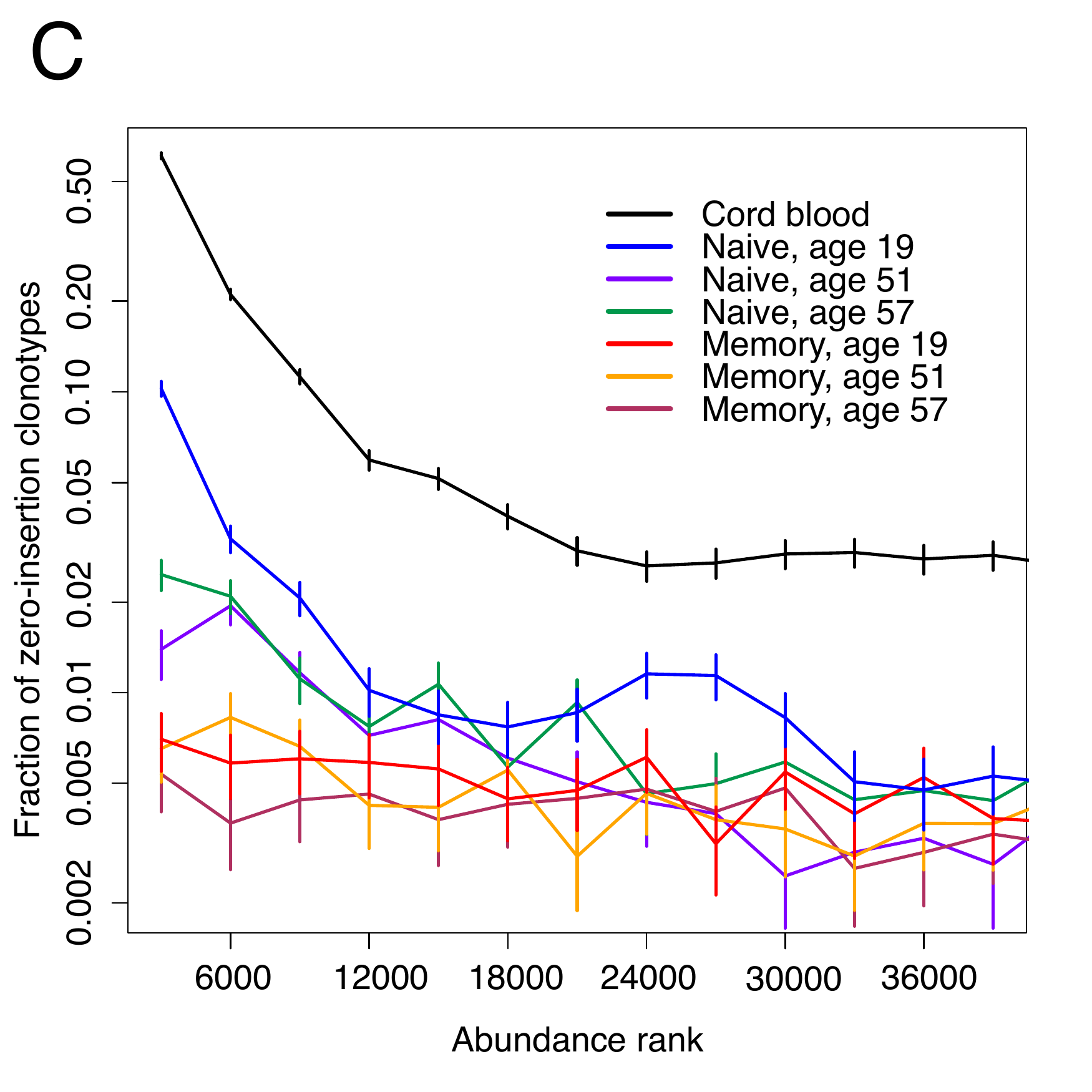}
\caption{{\bf The number of inserted nucleotides in in-frame TCR beta clonotypes depends on their abundance.} {\bf A}. Mean numbers of insertions were obtained by analysing groups of 3000 sequences of decreasing abundance. Clonotypes from the cord blood (black) show  a strong dependence on abundance, with high-abundance clones having much fewer insertions than low-abundance ones. Clonotypes in a young adult naive repertoire (blue) show a similar but less marked trend. Naive clonotypes in older adults (violet and green) show an even weaker trend. Adult memory samples of all ages show no dependence at all (red, yellow and maroon). Error bars show 2 standard errors. {\bf B}. Probability distributions for number of insertions for two rank classes for young naive and cord-blood samples (ranks 1-3000 on top, ranks 45000-48000 on bottom). For high-ranking sequences, the probability of having zero insertions is high both for adult naive and cord blood samples. For middle-ranking sequences, the probability of 0 insertions is much lower, and distributions are similar for adult naive and cord-blood samples. {\bf C}. Fraction of clonotypes with zero insertions for different abundance classes. Error bars show one standard deviation. 
}
\label{fig3}
\end{figure*}

To verify this hypothesis, we characterized the in-frame beta-chain repertoire of human cord blood (see SI Materials and methods).
One feature of the rearranged chains is
the number of insertions at the junctions between the gene segments (VD and DJ in the case of beta chains). We ranked beta TCR clonotypes from human cord blood data by decreasing abundances and plotted the mean number of insertions (inferred iteratively and averaged over groups of 3000 clonotypes, see SI Methods), as a function of this abundance rank (Fig.~\ref{fig3}A). The most abundant clones in cord blood had markedly smaller numbers of insertions (black line). 
The naive  repertoire of a young adult ({blue line}) showed a much weaker dependence on abundance than the cord blood repertoire, but followed a similar trend. {The dependence was even further reduced in older adults (purple and green lines).}
Interestingly, the number of insertions in the beta chains of the adult memory repertoire ({red, orange and maroon lines}) did not depend of the abundance of these cells. This observation can be explained by the resetting of the size of memory clones following an infection, erasing features of the abundance distribution inherited from fetal life.
Looking more closely into the distribution of the number of insertions (Fig.~\ref{fig3}B) reveals that low mean numbers of insertions are associated with an enrichment in clonotypes with zero insertions. Accordingly, the fraction of naive zero-insertion sequences generally decreased with abundance rank (Fig.~\ref{fig3}C), with again a stronger dependency in cord blood and young adults.
Fewer numbers of insertions in the cord blood are expected because TdT, the enzyme responsible for random insertions, is initially strongly downregulated in prenatal development \cite{Benedict2000}. This enrichment in low-insertion sequences persists and shows weak signatures in the adult naive repertoire, suggesting long lifetimes of cord blood clonotypes (although not necessarily of individual cells).

\begin{figure}
\noindent\includegraphics[width=\ff\linewidth]{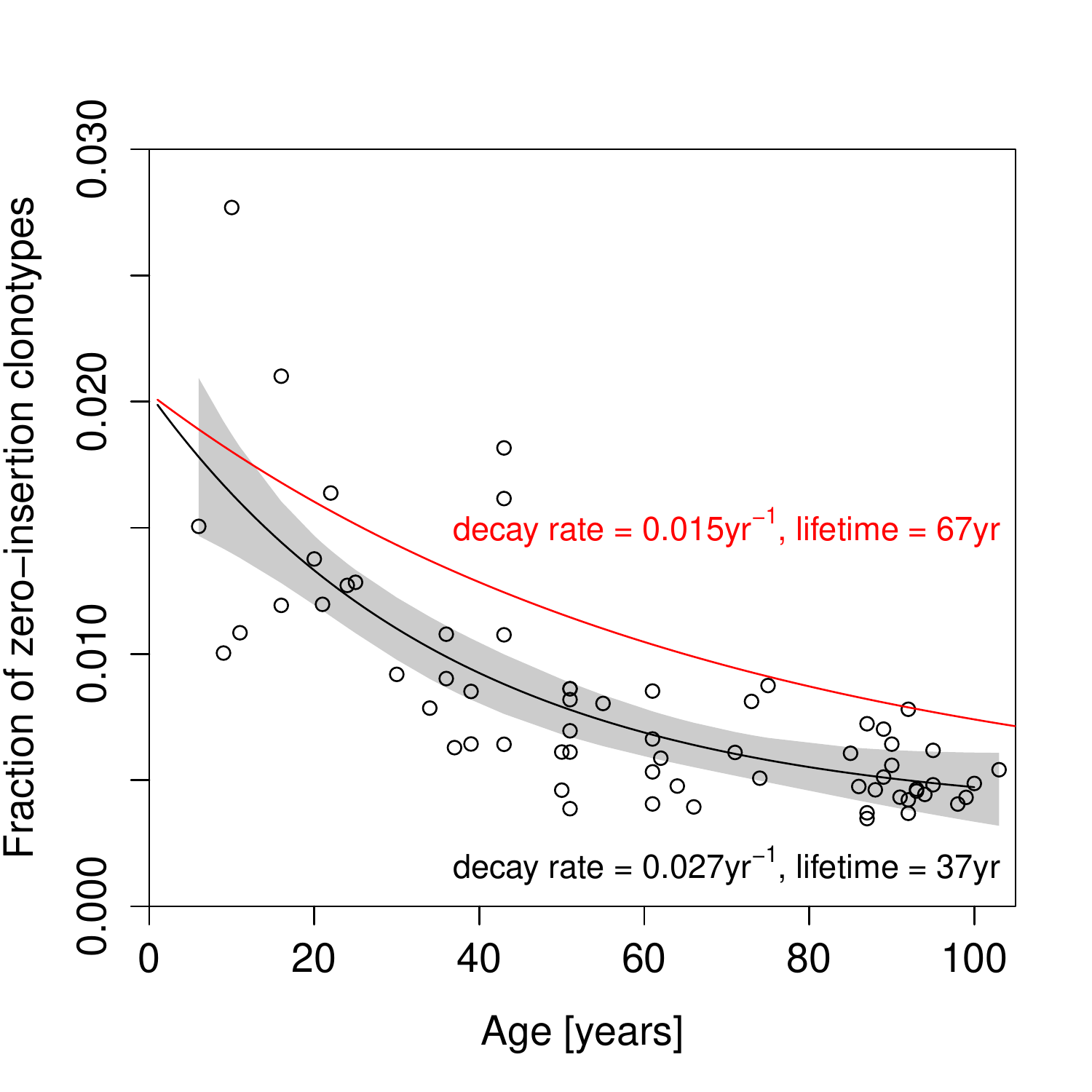}
\caption{{\bf Lifetime of abundant in-frame TCR beta clonotypes with zero insertions.} The fraction of zero-insertion clonotypes among the 2000 most abundant clonotypes as a function of age (circles) is well fitted by an exponentially decaying function of time (black curve). This decay is faster than would be predicted from the decay of the naive compartment alone (red curve), indicating a slow decay of zero-insertion clonotypes of fetal origin.}
\label{fig4}
\end{figure}

{The enrichment of zero-insertion sequences in large clonotypes 
can be used to verify the hypothesis of long lived fetal clonotypes originating from the cord blood. Analysing TCR beta repertoire data from individuals of different ages \cite{Britanova2014}, 
we observed a slow decay of abundant zero-insertion clonotypes with age, with a {characteristic time of {37} years} (Fig.~\ref{fig4}). However, the excess of abundant TdT- clonotypes of fetal origin only affects naive cells (Figs.~\ref{fig3} and S10), whose relative fraction is also known to decrease with time \cite{Britanova2014}.} 
{To assess the importance of this confounding effect, we fit an exponential decay model for the percentage of naive cells 
and found a characteristic decay time of {67} years. Therefore, the attrition of the naive pool alone cannot explain the decrease of zero-insertion clonotypes, which we attribute instead to the progressive extinction of clones of fetal origin, consistently with our hypothesis.} 

\begin{figure}
\noindent\includegraphics[width=\ff\linewidth]{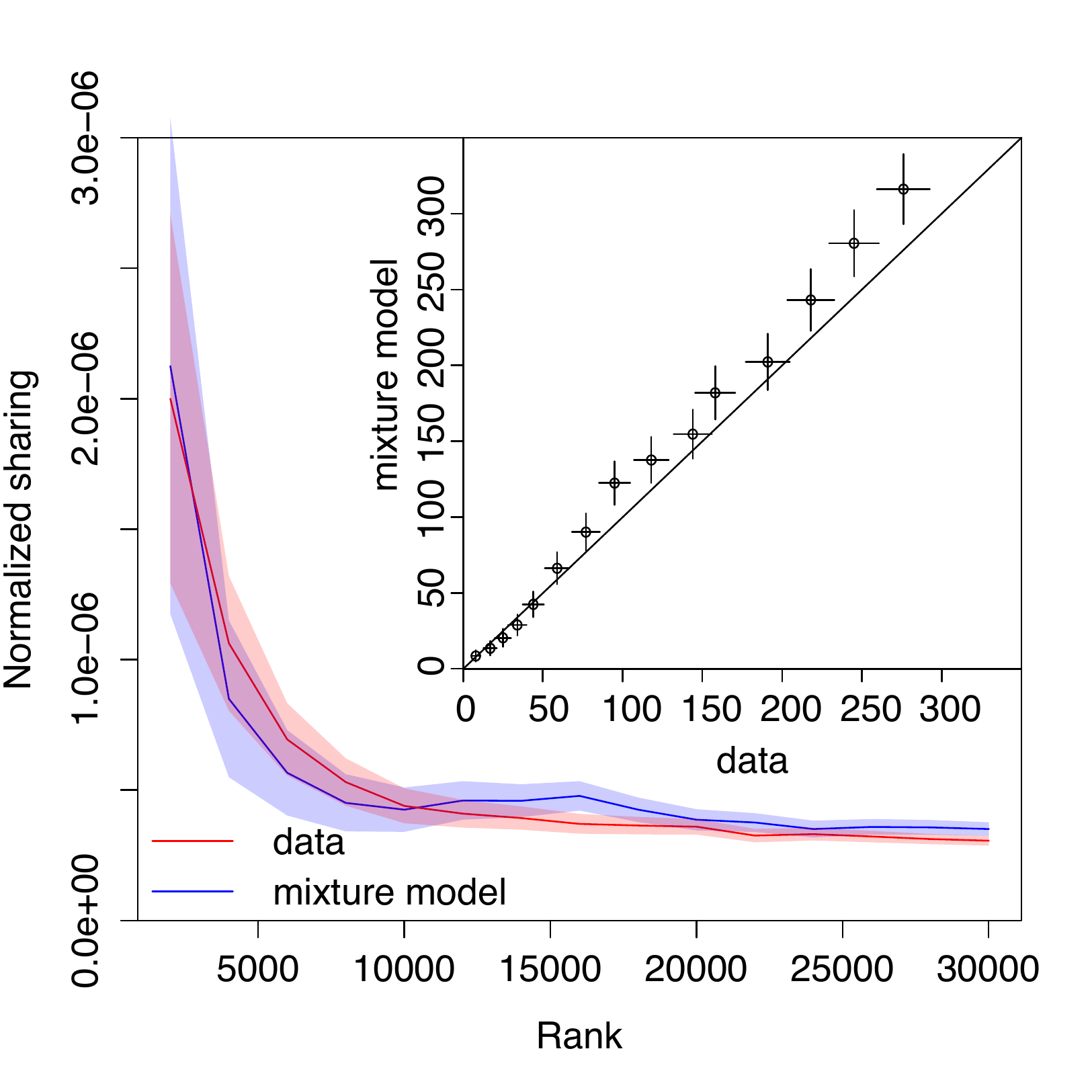}
\caption{{\bf Sharing of alpha out-of-frame TCR clonotypes as a function of clonal abundance.}
The normalized number of shared out-of-frame alpha CDR3 sequences between two individuals is showed as a function of clonotype abundance (e.g. normalized sharing for 2000 most abundant clones from both repertoires, 4000 most abundant, etc.), and compared to the amount of sharing that would be expected by chance (blue curve), taking into account the variable fraction of zero-insertion clonotypes as a function of their abundance. Data and predictions show excellent quantitative agreement (inset), with one fitting parameter. Error bars show one standard deviation.
\label{fig5}
}
\end{figure}

Fetal clones are directly exchanged between twins through cord blood and then persist with age, explaining the enhanced sharing between them.
At the same time, TdT- zero-insertion clonotypes are also most likely to be shared by chance than regular TdT+ sequences, because of their fewer number of insertions and reduced diversity. What are the implications of this observation for sharing between non-twins? Since zero-insertion sequences are overrepresented among abundant clonotypes (Fig.~\ref{fig3}), we predict that abundant {out-of-frame} clones are more likely to be shared.
To make our prediction quantitative, we built a mixture model of the out-of-frame alpha repertoire (see SI Text for details). We assumed that clonotypes of a given abundance $C$ are made up of a certain fraction $F(C)$ of TdT- zero-insertion clonotypes, and a complementary fraction $1-F(C)$ of regular TdT+ clonotypes. Because TdT+ clonotypes may also have no insertions, the fraction of the TdT+ and TdT- sets had to be learned in a self-consistent manner. Sequences were generated by the model for each abundance class $C$, and their sharing compared to the data.
The model accurately predicts the normalized sharing number of out-of-frame alpha-chain CDR3s as a function of clonotype abundance (Fig.~\ref{fig5}), up to the common multiplicative factor of 1.7 by which the non-mixture model generally underestimates CDR3 sharing (see Fig.~S2). 
Thus, the enhanced sharing of high-abundance clonotypes is entirely attributable to their higher propensity to have no insertions, making them more likely to be shared by chance.

Our results on the biological contamination of T cells in twins are robust to possible experimental artefacts.
First, our framework relies on the accurate counting of  TCR cDNA sequences using unique molecule identifiers\citep{Kivioja2011}. 
To exclude the possibility of contamination during the PCR and sequencing process, we double barcoded each cDNA library. To further exclude the possibility of early contamination of the blood samples, we performed replicate experiments at different times using different library preparation protocols. Comparison of repertoire overlaps from such replicate experiments for the same set of twins shows no difference and rules out experimental contamination as a confounding effect (see SI). We also observed the same effects in previously and independently collected datasets \cite{Zvyagin2014}, further excluding the possibility of experimental artefacts (Fig.~S8).

Cord blood sharing between twin embryos could have important implications on twin immunity: they could share and respond with private clonotypes, which would otherwise not be likely to be produced independently. This could possibly include sharing of malignant \cite{Teuffel2004,Ford1993,Wiemels1999} or autoimmune clones, leading to disease in both individuals. This mechanism of sequence sharing is very different from sharing by convergent recombination \cite{Venturi:2011co}, because it also implies the sharing of the second TCR chain and of the cell phenotype. Paired repertoires studies, which combine alpha and beta chains together \cite{Turchaninova2013c,Howie2015}, could be used to track clones shared between twins more precisely, and distinguish them from convergently recombined ones.

We conclude that fetal clonotypes are long-lived based on the analysis of over-abundant zero-insertion clonotypes. Invariant T-cells, MAIT (Mucosal-Associated Invariant T-cells) and iNKT (Invariant Natural Killer T-cells)  are intrinsically insertion-less, have restricted VJ usage for alpha chain, and are often abundant. These cells are produced in adulthood and could in principle constitute a substantial fraction of our zero-insertion dataset, confounding our analysis. Since our abundant zero-insertion clonotypes have a very diverse usage of VJ genes, we can exclude that the majority of them are from invariant T-cells, although we did identify a small number of such invariant TCR alpha chain clonotypes, see SI.

Our current data clearly shows that clonotypes that originated in the cord blood tend to be among the most abundant in the naive repertoire, but we cannot unambiguously point to the source of this effect. One possibility is convergent recombination \cite{Venturi:2011co}: high clonotypes abundances could be due to the accumulation of multiple convergent recombination events made more likely by the limited recombination diversity during fetal development.
An alternative explanation is that these clones have had more time to expand than others. Fetal cells come from different precursors, and mature in a different environment (the fetal liver), than post-natal cells \cite{Mold2010}. {\em In vitro} experiments have shown that fetal T-cells have a different proliferation potential than post-fetal cells \cite{Schonland2003}. Additionally, a vacant ecological niche effect may play a role. When these clones first appeared, the repertoire had not reached its carrying capacity set by homeostatic regulation, leaving room for future expansion. These clones may have initially filled the repertoire, later to be gradually replaced by post-fetal clonotypes. Consequently, fetal clones, including those whose TCR was recombined with no TdT, would be expected to have larger sizes. Clones with non-zero numbers of insertions could also originate in the cord blood as TdT is upregulated, however we have no way of distinguishing them from postnatal clones. 
Quantitative TCR repertoire profiling (preferably with the use of unique molecular identifiers for accurate data normalization and error correction), performed for species with no TdT activity in the embryo, such as mice, as well as novel cell lineage tracking techniques \cite{Naik2013} could be used to investigate the detailed dynamics of fetal clones.
This large initial expansion of fetal clones could protect them from later extinction.
This would suggest that the estimated {37}-year lifetime of zero-insertion fetal clonotypes could be longer than that of regular clones produced after birth.

The proposed large initial expansion of fetal clones could protect them from later extinction. This would suggest that the estimated 37-year lifetime of zero-insertion fetal clonotypes could be longer than that of regular clones produced after birth. Sharing of beta TCRs has previously been shown to decrease with age \cite{Britanova2014}. Depletion of fetal clonotypes, which are more likely to be shared, could contribute to this phenomenon. Our results also predict that the excess sharing of clonotypes between twins due to the biological contamination of fetal cells should decrease with age. In general, the observed abundance of large zero-insertion clonotypes and their persistence through significant part of our life should have important consequences for the adaptive immunity regulation both in pre- and post-fetal period.

Lastly, our general framework for analyzing the overlap between different repertoires has far-reaching practical implications for the tracking of T-cell clonotypes in the clinic. In particular, the analysis of overlap between pre- and post-treatment repertoires using probabilistic characteristics of clonotypes sharing could help determine the host or donor origin of clonotypes after hematopoietic stem cell transplantation (HSCT), and also increase reliability of malignant clones identification in minimal residual disease follow-up.

TRB/TRA libraries sequencing, raw sequencing data processing and reconstruction of TRB/TRA repertoires were supported by Russian Science Foundation grant №15-15-00178. Work was also partially supported by Russian Science Foundation project №14-14-00533 (to DMC, molecular barcoded data analysis). This work was partially supported by European Research Council Starting Grant n. 306312. We  want to thank Dr. I.V. Zvyagin for his help in finding appropriate donors for this study, and S.A. Kasatskaya for productive discussion of the manuscript. Part of experiments were conducted using equipment provided by the CKP IBCH RAS core facility.

\bibliographystyle{pnas}

\setcounter{figure}{0}
\setcounter{table}{0}
\renewcommand{\thefigure}{S\arabic{figure}}
\renewcommand{\thetable}{S\arabic{table}}

\section{Supplementary materials and methods}

\subsection{Blood samples}

Blood samples were collected from 3 pairs of monozygotic twin female donors, 23 (donors S1 and S2), 23 (donors P1 and P2) and 25 (donors Q1 and Q2) years old respectively. We also collected blood from two 19 and 57 year old male donors, along with a 51 year old female donor for memory and naive T-cells isolation, and a cord blood sample from a female newborn. All donors were healthy Caucasians, blood samples were collected with informed consent, and local ethical committee approval. The genetic identity of the twins was checked using polymorphic Alu insertion genotyping \cite{Mamedov2010}.

PBMCs were isolated from 12 ml of blood using Ficoll-Paque (Paneco, Russia) density gradient centrifugation. One third of the isolated PBMCs was used for total RNA isolation with the Trizol reagent (Invitrogen, USA) according to the manufacturer's protocol. Other cells were used for CD4, CD8 and CD45RO+ T-cells isolation. 

\subsection{CD4, CD8, 45RO+ T-cell isolation}

CD4 and CD8 T-cells were isolated from PBMCs using the CD4+ and CD8+ positive selection kit (Invitrogen, USA) according to the manufacturer's protocol. CD8 T-cells were isolated from CD4 depleted samples to maximize the cell yield. 45RO+ cells were extracted using human CD45RO microbeads (Myltenyi, USA). Naive T-cells were isolated with the CD8+ T-cell naive isolation kit (Myltenyi, USA) according to the manufacturer's protocol without the final CD8 enrichment step. 

Total RNA was immediately extracted from the isolated cells using the Trizol reagent (Invitrogen).

\subsection{TCR $\alpha$ and TCR $\beta$ cDNA library preparation}
\begin{figure}[p]
\noindent\includegraphics[width=\linewidth]{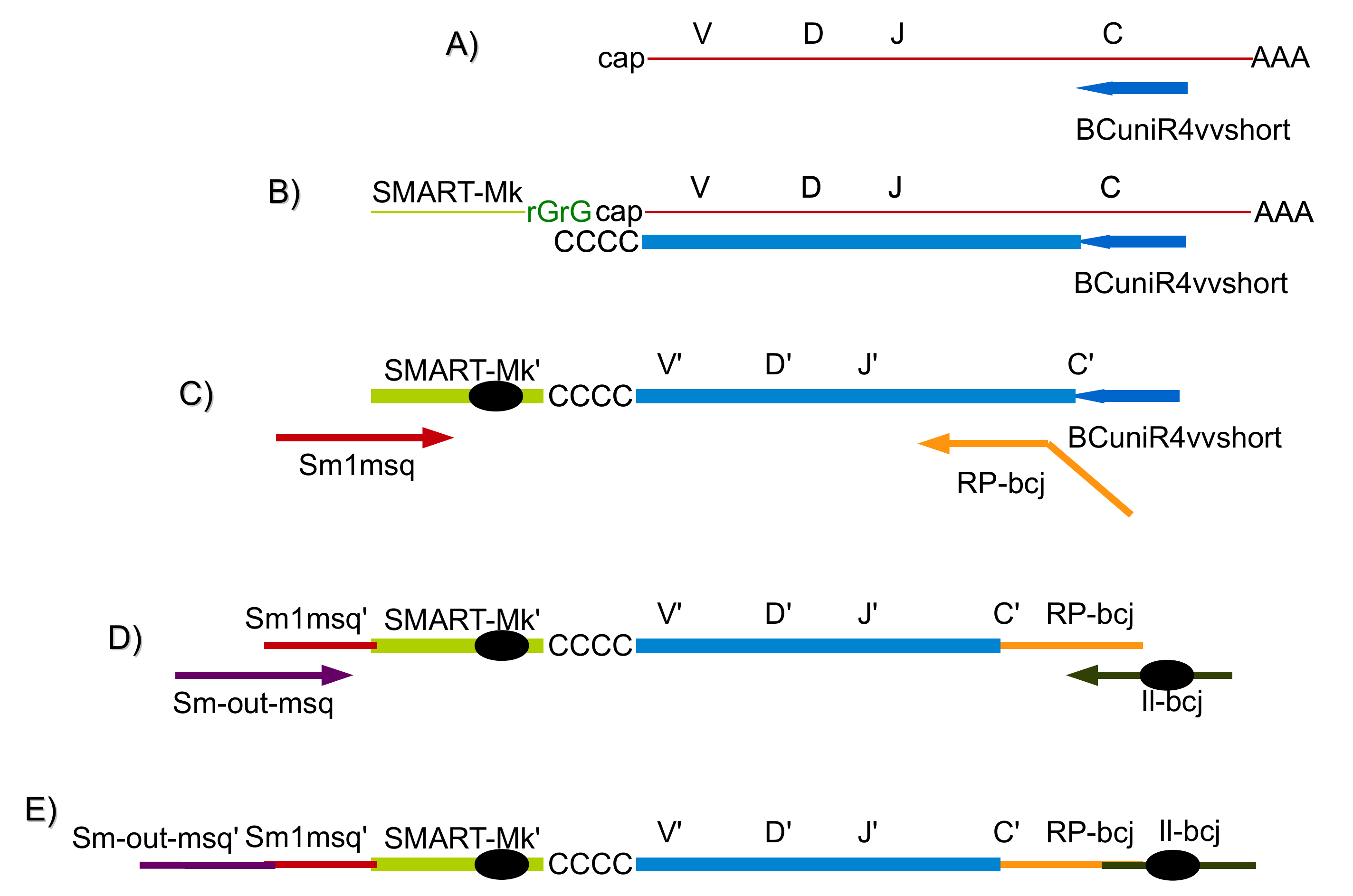}
\caption{{\bf Library preparation protocol.} A) cDNA first strand synthesis for alpha and beta chains starts from specific primers in the C-segment conserved region. B) The template switching effect was used to introduce a universal primer binding site to the 3'cDNA end. The SMART-Mk sequence contains a sample barcode (black ellipse) for contamination control. C) and D) In two subsequent PCR steps we introduce the TruSeq adapter sequences along with Illumina sample barcodes (black ellipse). E) The resulting cDNA molecule is double barcoded, contains a Unique Molecular Identifier (UMI) and is suitable for direct sequencing on the Illumina HiSeq platform with the custom primers.}
\label{figS1}
\end{figure}

The library preparation protocol was adapted from \cite{Mamedov2013} with modifications. 
The cDNA first strand was produced from the total RNA using the SmartScribe kit (Clontech, USA) and universal primers specific for the C-segment (see Fig.~\ref{figS1} A). Custom cap-switching oligonucleotides with unique molecular identifiers (UMI) and sample barcodes were used to introduce the universal primer binding site to the 3' end of the cDNA molecules (see Fig.~\ref{figS1} B). Each tube contained 500 ng of total RNA, 1x SmartScribe buffer, dNTP (1 mM each), 10pcmol of BCuniR4vvshort and TRACR2 primers (see Table S1 for sequences) and 1 $\mu$l of SmartScribe reverse transcriptase. 5mkg of the total RNA was used for the cDNA synthesis for each sample (10 tubes per sample). The cDNA synthesis product was treated (45 min, 37$\degree$C) with 1 $\mu$l of 5u/$\mu$l UDG (NEB, USA) to digest the cap-switching oligonucleotide and purified with the Quigen PCR purification kit. After the cDNA synthesis two steps of PCR amplification were used to amplify the cDNA and also introduce Illumina TruSeq adapters as well as the second sample barcode. After both steps the PCR product was purified using the Quigen PCR purification kit according to the manufacturer's protocol. The first PCR step (see Fig.~\ref{figS1} C) consists of 16 cycles of: 94 $\degree$C for 20 sec, 60$\degree$C for 15 sec, 72$\degree$C for 60 sec. Each tube contained (total reaction volume 15 $\mu$l) 1x Q5 polymerase buffer (NEB), 5 pmol of Sm1msq and RPbcj1, RPbcj2, RPacj primers, dNTP(0.125 mM each) and 0.15 $\mu$l of Q5 polymerase.  Then 1 $\mu$l of the purified PCR product was used for the second amplification step (see Fig.~\ref{figS1} D) consisting of 12 cycles of: 94$\degree$C 20 sec, 60$\degree$C 15 sec, 72$\degree$C 40 sec. Each tube contained (total reaction volume 25 $\mu$l): 1x Q5 polymerase buffer, 5 pmol of Smoutmsq and Il-bcj-ind or Il-acj-ind primers (with sample specific indices, for beta and alpha libraries respectively, one primer per sample), dNTP(0.125 mM each) and 0.25 $\mu$l of Q5 polymerase. Size selection for 500-800bp fragments of the purified PCR product was performed using electrophoresis in 1\% agarose gel. 

\subsection{Next Generation Sequencing}

cDNA libraries were sequenced on the Illumina HiSeq platform (2x100nt). Custom sequencing primer sequences are listed in Table S1. The total numbers of sequencing reads are shown in Table S2.

\subsection{Raw data preprocessing}
Raw sequencing data files were preprocessed with MiGEC \cite{Shugay2014}, sequencing reads were clustered by unique molecular identifiers (UMI). UMIs with less than two reads were discarded to reduce the number of erroneous sequences. Then sequences were processed with MiXCR \cite{Bolotin2015} to determine the CDR3 position and nucleotide sequence. For the  numbers of UMIs after filtering see Table S2.
 
\subsection{Learning recombination statistics}
We built a generative model that describes the probability of generation of recombined sequences, following the theoretical framework described in \cite{Murugan2012, Elhanati2015b}. The generation probability for each sequence is calculated as the sum over all recombination scenarios $r$ that can produce that sequence, $P_\text{gen}(\text{sequence})=\sum_r{P_\text{rearr}(r)}$.
For TCR alpha chains the model assumes the following factorized form for a recombination scenario defined by the choice of genes (V and J), $P(V,J)$, deletions (delV and delJ), $P(\text{del}V|V)$ and $P(\text{del}J|J)$ and insertions (ins), $P({\rm ins})$:
\begin{equation}
P_\text{rearr}^\alpha(r) = P(V,J) P(\text{del}V|V) P(\text{del}J|J) P({\rm ins}). 
\label{P_alpha}
\end{equation}

The parameters of the models, the different probabilities in the factorized formula, were inferred by maximizing the likelihood of the observed out-of-frame sequences given the model, using Expectation-Maximization \cite{Murugan2012}. 
For alpha chains, the model was reformulated as a Hidden Markov Model, and the parameters were learned efficiently using a Baum-Welch algorithm, as described in \cite{Elhanati2015b}.

For beta chains, the model describes probabilities for V, D and J choices, with possible deletions and insertions at each of the two junctions:

\begin{eqnarray}
P_\text{rearr}^\beta (r) &=& P(V,D,J) P(\text{del}V|V) P(\text{ins}VD)   \\
&& \times P(\text{del}Dl,\text{del}Dr|D) P(\text{ins}DJ) P(\text{del}J|J) \nonumber
\end{eqnarray}

The parameters for the beta chain model were inferred directly using the Expectation-Maximization algorithm, by enumerating all possible recombination scenarios that can produce each sequence, using the procedure described in \cite{Murugan2012}.
 
This procedure allows us to learn the features of the recombination statistics with great accuracy, in particular the distribution of number of insertions at the junctions, even though the recombination events themselves cannot be unambiguously be determined for each sequence because of convergent recombination.

\subsection{Distribution of insertions for each beta chains abundance class} 
We applied the procedure described in the previous section separately for each abundance class of the beta-chain sequences.
However, given the small size of the datasets  (2000 or 3000 sequences), we did not learn the full model for each class. Instead, we used a previously inferred universal beta-chain recombination model \cite{Murugan2012} for the V,D,J gene usages and their deletion profiles, and we learned the insertion distributions ($P(\text{ins}VD)$ and $P(\text{ins}DJ)$) for each class separately, while keeping the other parameters constant.
The distribution of insertions thus inferred are used to plot the results of Figs.~3 and 4 of the main text.

\subsection{Inference of selection factors}
In-frame sequences statistically differ from out-of-frame sequences (besides their frameshift), because in-frame sequences are functional and have passed thymic selection.
For each sequence we defined a selection factor  $Q$ as the ratio of the probability of observing the sequence in the in-frame set, to the probability of recombining the sequence according to out-of-frame statistics (as inferred above).  The overal selection factor $Q$ is assumed to be the product of several independent factors $q$ acting on the CDR3 length $L$ and
on the identity of amino acid $a_i$ at each position $i$ of the CDR3 \cite{Elhanati2014}:
\begin{equation}
{Q}\propto \ q_L \   \prod_{i=1}^L q_{i;L}(a_i)
\end{equation}
The parameters were inferred by maximizing the likelihood with gradient ascent, as described in \cite{Elhanati2014}. 
 
\subsection{Data analysis}
Analysis of the shared clonotypes was performed using the R statistical programming language \cite{R} and the tcR package \cite{Nazarov2015}.
 
\subsection{Out-of-frame sharing prediction}
To predict sharing for each individual, we generated sequences using our recombination model $P_{\rm gen}$ (alpha or beta), with individually inferred model parameters. Normalized sharing of the TCR sequences between two clonesets is defined as the number of the same unique TCR nucleotide sequences observed in both of them, divided by the product of the total numbers of unique TCR nucleotide sequences in the two datasets. 

We calculated sharing of either whole chains, or of their CDR3, defined as the sub-sequence going from the conserved cystein at the end of the V region, to the conserved phenylalanine in the J region.

\begin{figure}[p]
\noindent\includegraphics[width=\linewidth]{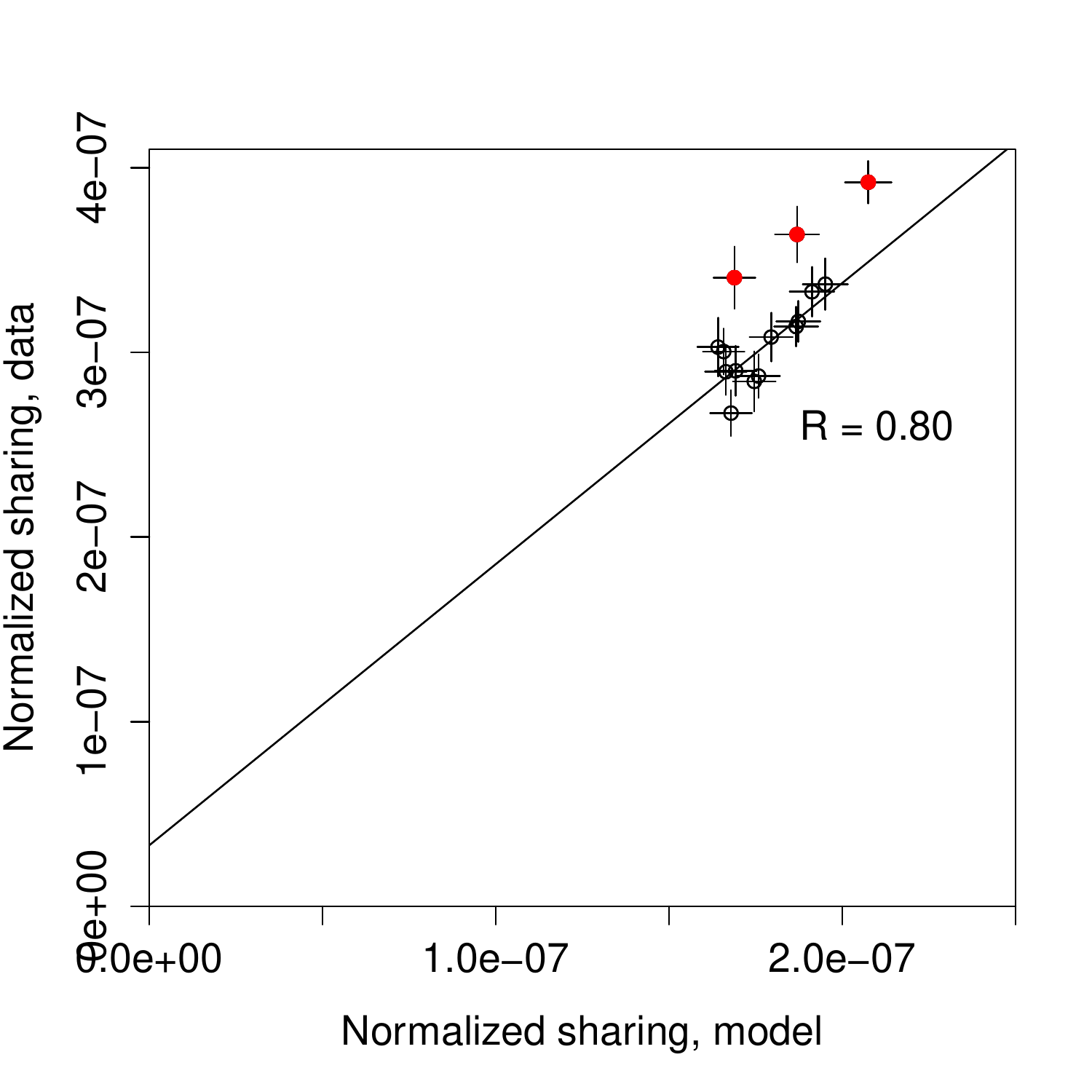}
\caption{{Number of shared out-of-frame alpha TCR CDR3 clonotypes reported between all $15$ pairs of $6$ donors consisting of $3$ twin pairs (ordinate) compared to the model prediction (abscissa). To be able to compare datasets of different sizes, the sharing number was normalized by the product of the two cloneset sizes. The outlying three red circles represent the twin pairs, while the black circles refer to pairs of unrelated individuals. Error bars show one standard deviation. The diagonal line is a linear fit for unrelated individuals, of slope 1.7.}}
\label{figS10}
\end{figure}

The alpha chain results for whole-chain sharing are plotted in the main text in Fig. 1, and the data shows good agreement with the model. The results for CDR3 sharing are shown in Fig.~\ref{figS10}. 
The model systematically underestimates the normalized sharing by a common multiplicative factor of 1.7 for non-twins, with a Pearson correlation coefficient of  0.8 
between the data and the model prediction. Absolute numbers of shared CDR3 sequences for alpha chains varied from 400 to 1200.

\begin{figure}[p]
\noindent\includegraphics[width=\linewidth]{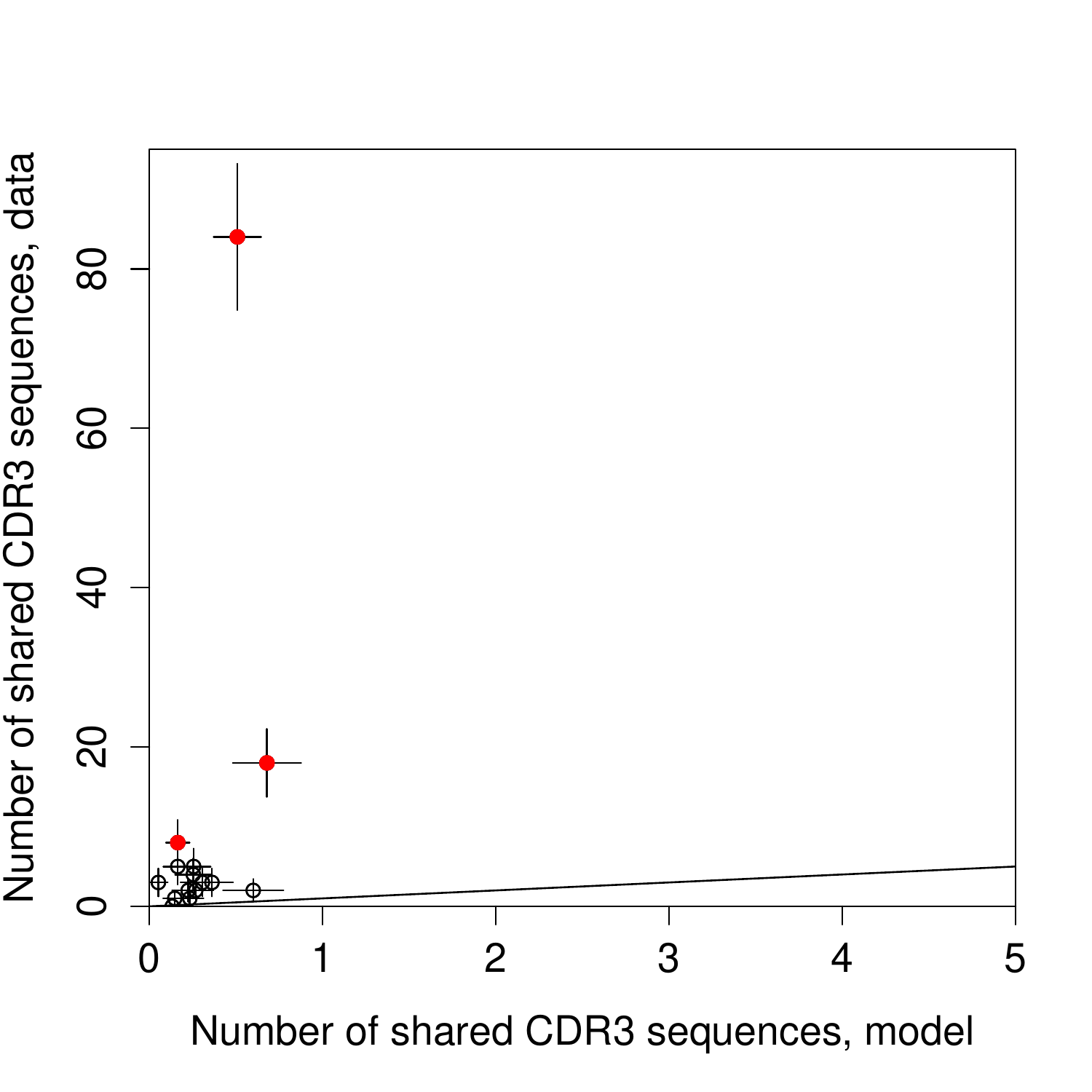}
\caption{{Number of shared out-frame beta TCR CDR3 clonotypes reported between all $15$ pairs of $6$ donors consisting of $3$ twin pairs (ordinate) compared to the model prediction (abscissa). The three outlying red circles represent the twin pairs, while the black circles refer to pairs of unrelated individuals. Error bars show one standard deviation.}}
\label{figS6}
\end{figure}

For beta chain sequences, the prediction of out-of-frame sharing is more difficult because of the low numbers of out-of-frame sequences in the RNA data, which, combined to a lower mean $P_{\rm gen}$, results in a much lower number of shared out-of-frame sequences. We also identified and removed from the dataset $26$ out-of-frame sequences shared between more than two individuals. These sequences are likely to arise due to reproducible aligner errors or technology artifacts -- some of them contained intronic sequences, etc. Absolute numbers of shared beta CDR3 sequences varied from 0 to 82.
Nevertheless, the number of shared beta out-of-frame CDR3 sequences for twins exceeded the model prediction (see Fig.~\ref{figS6}), confirming our hypothesis of biological contamination during pregnancy. 

\subsection{In-frame sharing prediction}
To accurately predict the normalized sharing number for in-frame nucleotide clonotypes, we generated sequences from $P_{\rm gen}$ as we did for out-of-frame sequences, but
weighted them by their selection factor $Q$ to account for thymic selection. The predicted normalized sharing number was then calculated as:
\begin{equation}
\frac{1}{|S_1|\cdot|S_2|}\sum_{s\in S_1 \cap S_2} Q^{(1)}(s)Q^{(2)}(s),
\end{equation}
where $S_1$, and $S_2$ are two synthetic sequence samples drawn from two models $P^{(1)}_{\rm gen}$,$P^{(2)}_{\rm gen}$ individually learned from the out-of-frame sequences of two individuals, and $Q^{(1)}(s)$, $Q^{(2)}(s)$ are selection factors learned individually from these two individuals' in-frame sequences. $|S_1|$ and $|S_2|$ denote the size of the two samples.
The sum runs over sequences $s$ found in both samples.

\begin{figure}[p]
\noindent\includegraphics[width=\linewidth]{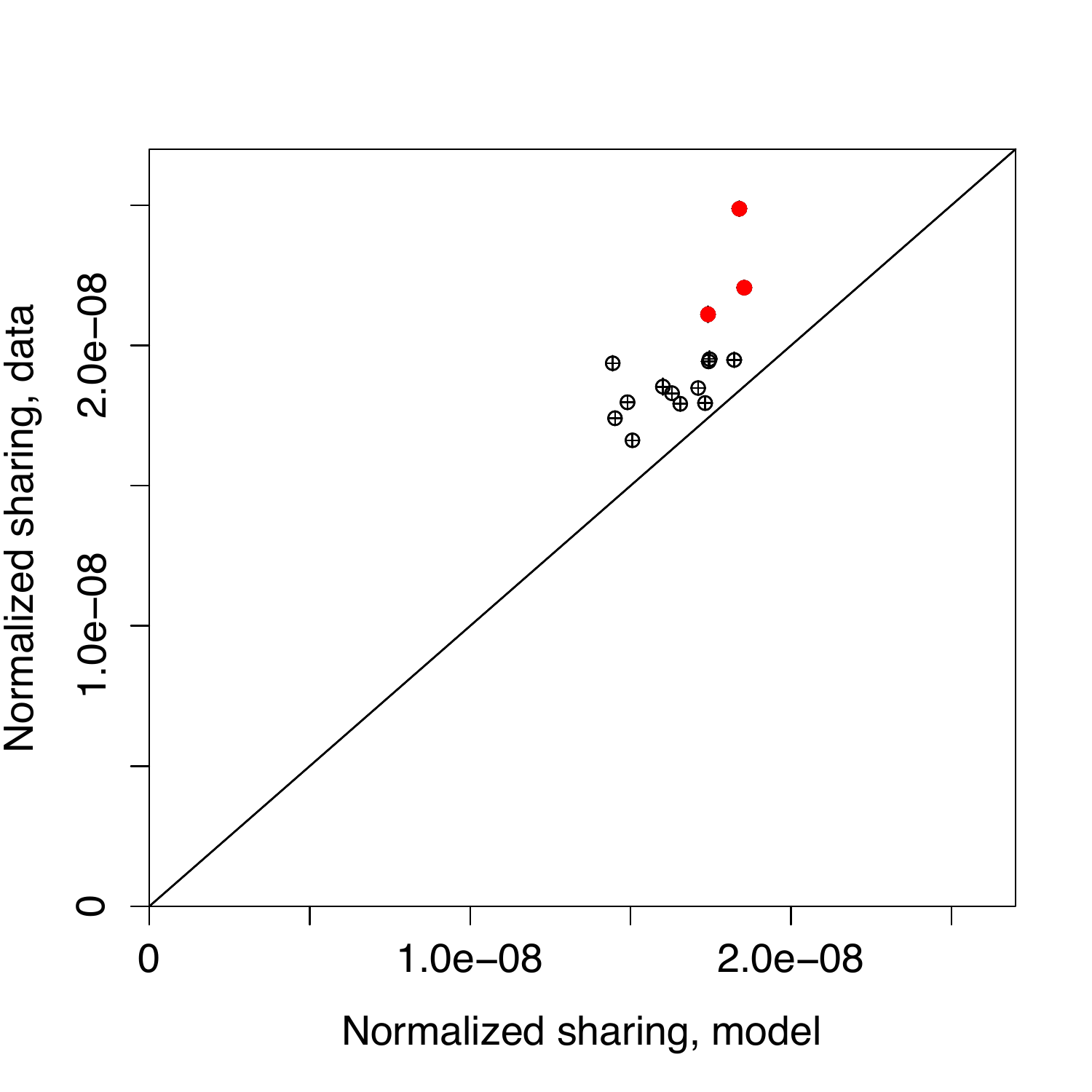}
\caption{{Number of shared in-frame beta TCR CDR3 clonotypes reported between all $15$ pairs of $6$ donors consisting of $3$ twin pairs (ordinate) compared to the model prediction (abscissa). To be able to compare datasets of different sizes, the sharing number was normalized by the product of the two cloneset sizes. The three outlying red circles represent the twin pairs, while the black circles refer to pairs of unrelated individuals. Diagonal is equality line. Error bars show one standard deviation.}}
\label{figS5}
\end{figure}

\begin{figure}[p]
\noindent\includegraphics[width=\linewidth]{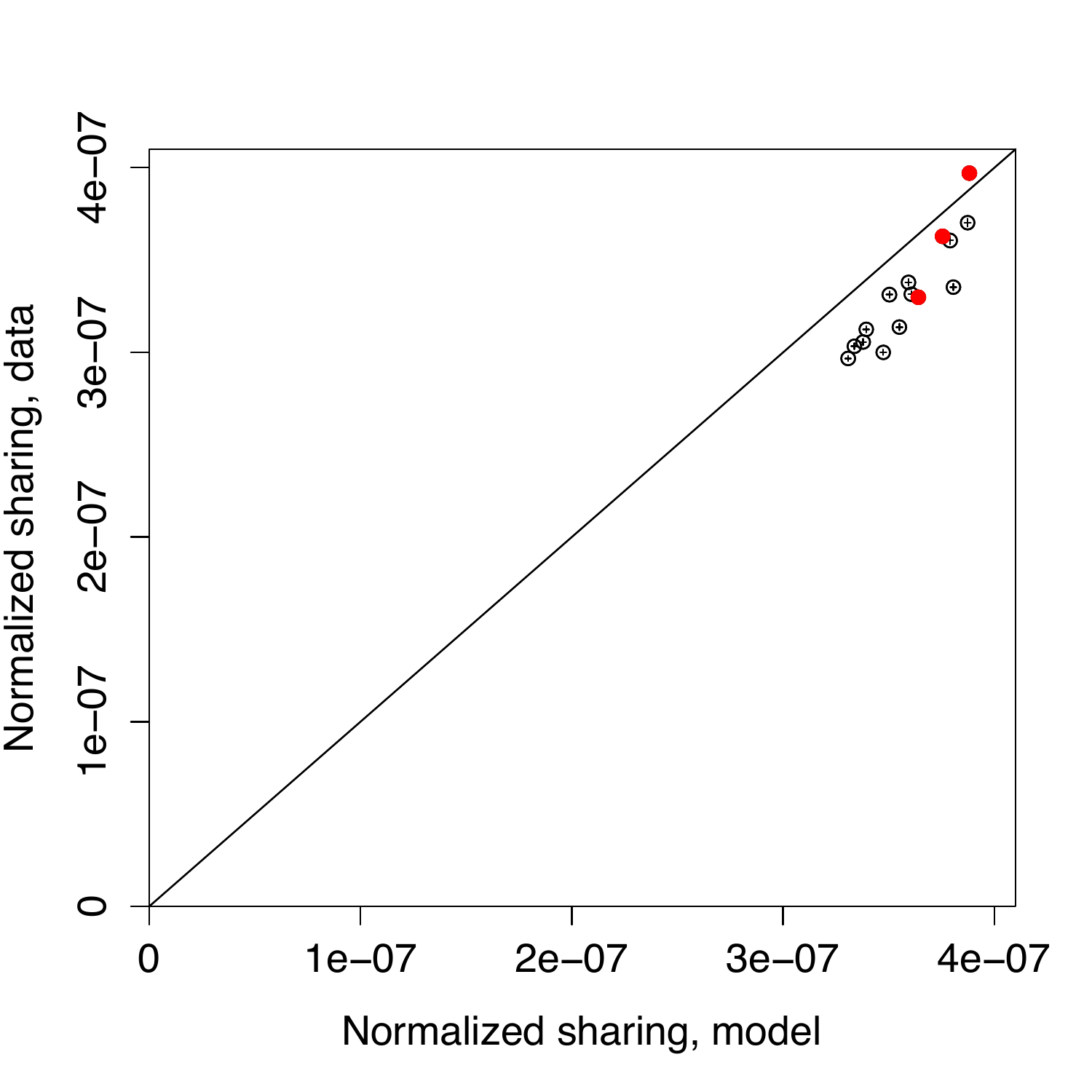}
\caption{{Number of shared in-frame alpha TCR CDR3 clonotypes reported between all $15$ pairs of $6$ donors consisting of $3$ twin pairs (ordinate) compared to the model prediction (abscissa). To be able to compare datasets of different sizes, the sharing number was normalized by the product of the two cloneset sizes. The three red circles represent the twin pairs, while the black circles refer to pairs of unrelated individuals. Diagonal is equality line.}}
\label{figS4}
\end{figure}

For both the beta and the alpha chains, the prediction agrees very well with the data (Fig.~\ref{figS5} and Fig.~\ref{figS4}). For the beta chain, twins share more CDR3 sequences than non-twin pairs, while no such effect was observed for the alpha chain sequences. This fact could be explained by the much higher number of clonotypes shared due to convergent recombination in the alpha in-frame dataset than in the beta in-frame and alpha and beta out-of-frame datasets. Excess of shared CDR3 nucleotide sequences due to biological contamination in twins is lower than the amount of convergent recombination noise in the alpha in-frame shared CDR3 nucleotide sequences. Absolute numbers of shared in-frame CDR3 sequences for alpha chains varied from 30000-50000 sequences depending of the pair, and 5000-9000 for beta chains. 

\subsection{Mixed model inference}
We hypothesized that the larger amount of zero insertion clonotypes is responsible for the increase in sharing between the most abundant clonotypes of the out-of-frame repertoires of unrelated individuals. To test this hypothesis, we constructed a mixture model for each abundance class, each class containing 2000 clonotypes ranked by decreasing abundance.

We assume that abundance class $C$ contains a fraction $F(C)$ of clonotypes generated with zero insertions, and $1-F(C)$ of regular clonotypes. Obtaining $F(C)$ is not straightforward because regular clonotypes can also zero insertions. In addition, the number of insertions cannot be determined with certainty -- for example, a deletion followed by an insertion matching the germline sequence can be wrongly interpreted as a case of no insertions.

To circumvent this problem, we determine for each abundance class a simpler quantity to estimate, namely the fraction $F_0(C)$ of clonotypes that are consistent with zero insertions, {\em i.e.} that can be entirely matched to the germline genes. Because of the reasons outlined above, $F_0(C)$ is {\em not} equal to $F(C)$. However, $F_0(C)$ is a linear function of $F(C)$, $F_0(C)=A+BF(C)$. Therefore, if we can generate synthetic sequences such that their $F_0(C)$ agrees with data, then we are guaranteed that their $F(C)$ will coincide with the data as well, even if we do not know the explicit mixing parameters $F(C)$.

To obtain this mixture, we generated many sequences from our recombination model $P_{\rm gen}$. To determine which generated sequences were consistent with zero insertions, we aligned them to all possible V and J genomic templates. We then separated out the sequences consistent with zero insertions from the others, and created, for each abundance class $C$, and artificial dataset with a fraction $F_0(C)$ of such sequences, and $1-F_0(C)$ of the other sequences (not consistent with zero insertions), where $F_0(C)$ is given by the data.

We then calculated normalized sharing in the synthetic data by including an increasing number of abundance classes, starting with the most abundant ones, and compared to data in Fig.~5.

\subsection{Exponential decay regression}
We fit an exponential decay curve to the fraction $Z$ of zero-insertion clonotypes in the 2000 most abundant clones as a function of age $T$ (Fig.~4):
\begin{equation}
Z\approx c + a  \exp(-b  T).
\end{equation}
We found $c=0.00363\pm 0.00154$, $b=0.0272 \pm 0.0091\ \mathrm{yr}^{-1}$, and $a=0.016696 \pm=0.00188$ using the nlm2 R package.

We fitted an analogous model for the attrition of the naive T-cell pool, {\em i.e.} the fraction $N$ of naive T-cells as identified using flow cytometry (see \cite{Britanova2014} for details):
\begin{equation}
 N \approx  a'  \exp(-b' T).
 \end{equation}
We obtained $a'=0.68\pm 0.054$ and $b'=0.01485\pm 0.0018\ \mathrm{yr}^{-1}$.

The data used in the two model fits are available in Table S3.

\section{Supplementary results}

\subsection{Distinctive properties of shared clonotypes between twins}

\begin{figure}[p]
\noindent\includegraphics[width=\linewidth]{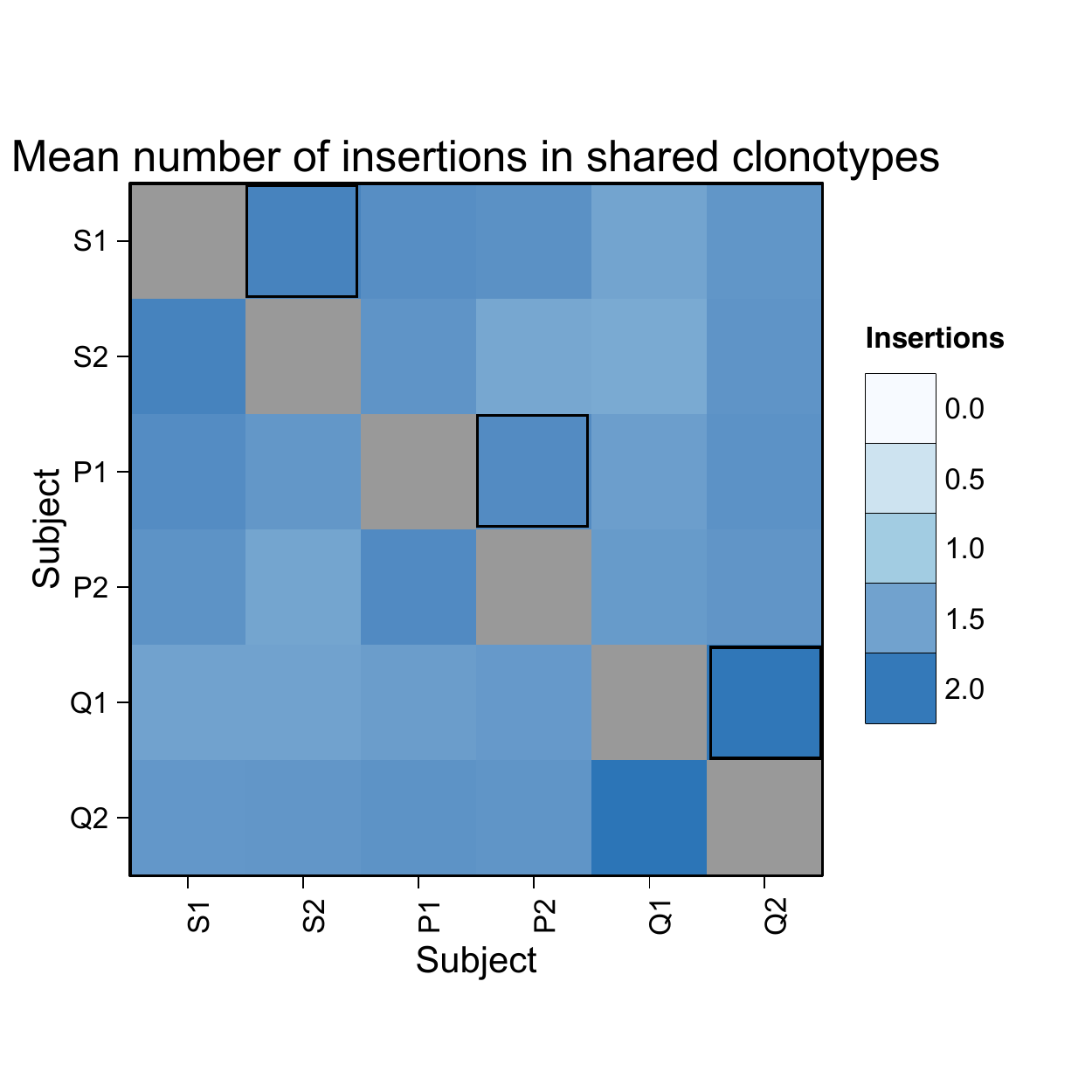}
\caption{{Mean number of insertions in shared sequences in alpha out-of-frame repertoires.}}
\label{figS2}
\end{figure}

\begin{figure}[p]
\noindent\includegraphics[width=\linewidth]{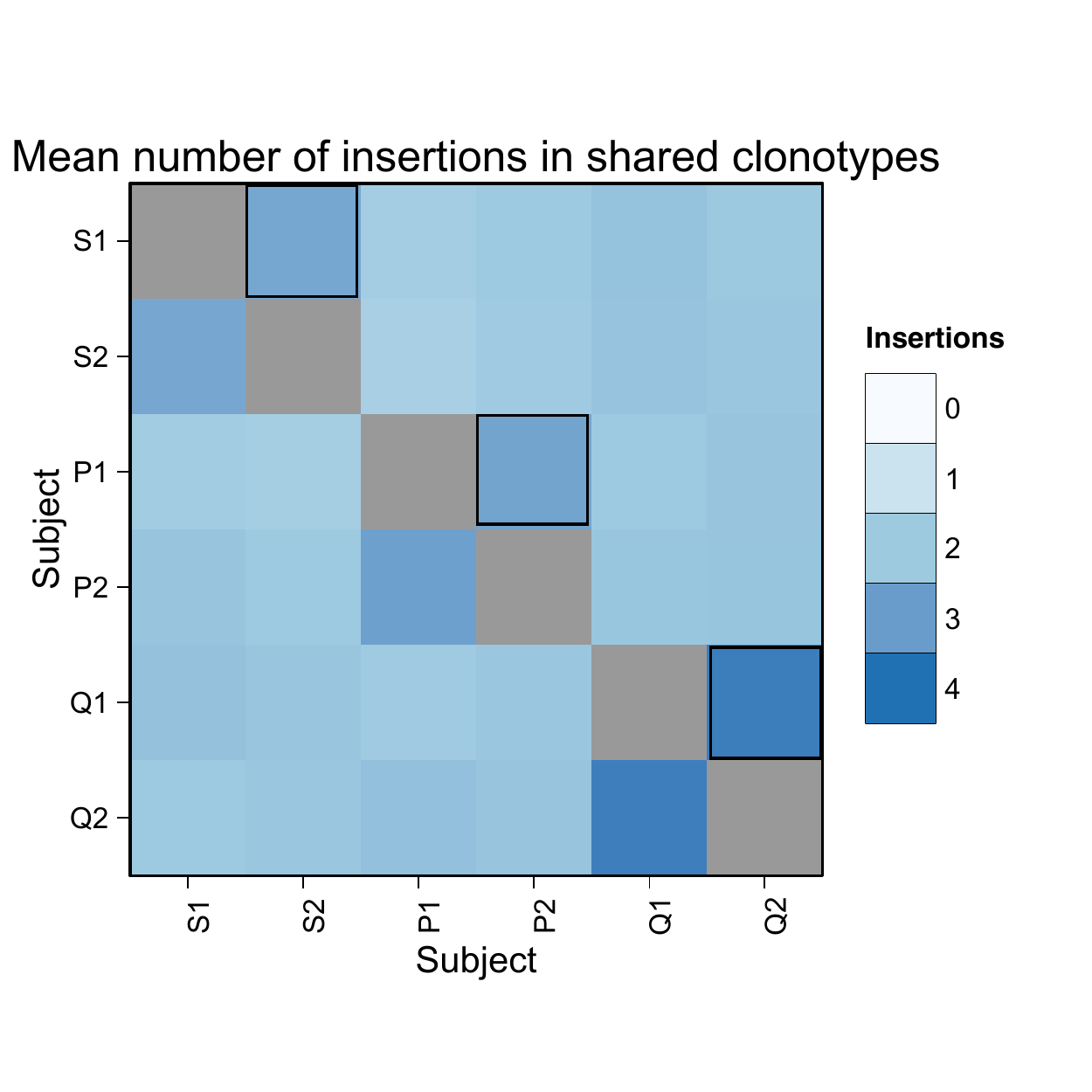}
\caption{{Mean number of insertions in shared sequences in alpha out-of-frame repertoires of CD45RO+ (memory) cells.}}
\label{figS3}
\end{figure}

Shared clonotypes in unrelated individuals appear in the process of convergent recombination. Sequences with a higher $P_{\rm gen}$ are thus more likely to be shared, and we can calculate accurately the distribution of $P_{\rm gen}$ among shared sequences (see Fig.~2).
We observe that sequences shared between twins violate this prediction, consistent with our hypothesis that some of these sequences are due to biological contamination.
To confirm this, we used a sequence feature that is negatively correlated with $P_{\rm gen}$ \cite{Murugan2012}: the number of insertions in the CDR3 region. The number of insertions in CDR3 sequences shared between unrelated individuals was indeed lower (Fig.~\ref{figS2}) than the mean number of insertions in non-shared sequences. However, the mean number of insertions in sequences shared between twins (black boxes) is higher than in unrelated individuals. The same and even stronger effect is observed for memory (CD45RO+) cells (Fig.~\ref{figS3}).

\subsection{The phenotype of beta chain out-of-frame shared clonotypes}
Two individuals displayed the most prominent excess of shared beta out-of-frame sequences. Since the model prediction for the number of shared sequences is close to zero we suppose that most of these shared sequences did not arise due to convergent recombination. These out-of-frame clones bear a second functional allele (otherwise they would have been filtered by selection in a thymus), and they also should have either the CD4 or the CD8 phenotype. To attribute these clonotypes a phenotype we separately sequenced CD4, CD8 and CD45RO positive subsets for the two donors and searched for the $84$ out-of-frame CDR3s shared between the unpartitioned out-of-frame repertoires. $44$ CDR3s were found in the CD8 subsets of both individuals, and only $5$ sequences were found in the CD4 subsets of both individuals. $25$ out of the $44$ CD8 and $3$ out of the $5$ sequences were also found in the 45RO+ compartment. Only $3$ sequences were mapped discordantly (e.g. CD4 in  one twin and CD8 in the second twin), and $2$ sequences were absent from the CD4, CD8 and CD45RO compartments of both individuals. For the other $32$ sequences the CD4/CD8 status could be determined only for one individual (most probably due to the sequencing depth limitations). In case of convergent recombination it is unlikely that shared nonproductive sequences would have the same phenotype in different donors. The phenotypic study thus confirms the biological contamination hypothesis. 

\subsection{Our results are reproducible using previously published data}
\begin{figure}[p]
\noindent\includegraphics[width=\linewidth]{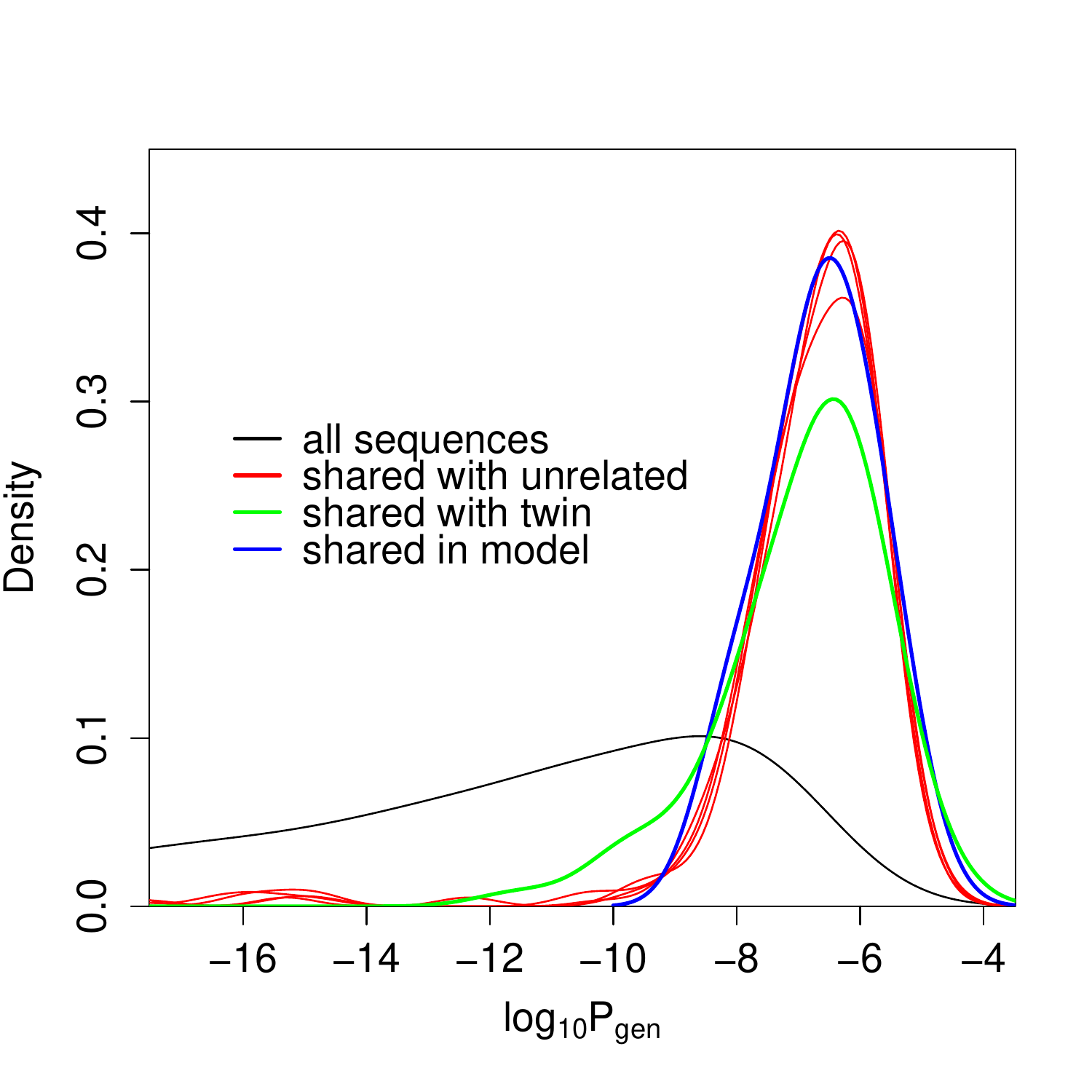}
\caption{{\bf Reproducibility of our results using previously published data.} Distribution of  $P_{\rm gen}$ -- the probability that a sequence is generated by the VJ recombination process -- for shared out-of-frame TCR alpha clonotypes between individual $A_1$ from \cite{Zvyagin2014} and the other five individuals. While the distribution of shared sequences between unrelated individuals (red curves) is well explained by coincidental convergent recombination as predicted by our stochastic model (blue curve), sequences shared between two twins (green curve) have an excess of low probability sequences. For comparison the distribution of $P_{\rm gen}$ in regular (not necessarily shared) sequences is shown in black.}
\label{figS7}
\end{figure}

We tested the robustness of our results on previously published twin data from \cite{Zvyagin2014}. 
We observed the same excess of low-probability shared sequences in twins compared to unrelated individuals as in Fig.~2
(see Fig.~\ref{figS7}).
These data also allowed us to control for possible experimental contamination. One of the twin pairs that participated in the present study was sequenced three years ago, using a different technology described in \cite{Zvyagin2014}, excluding the possibility of any contamination  between the old and new samples. Out of $84$ beta out-of-frame clonotypes shared between two new twin samples, $59$ were also shared between the new sample of one twin, and the old sample of the second twin. Therefore the out-of-frame sequences shared between the twins are reproducible and could not be result of experimental contamination with PCR-products or RNA.

\subsection{Invariant T-cell alpha clonotypes in the data}
It was previously shown that mucosal-associated invariant T-cells (MAIT) and natural killer T-cells (NKT) have an invariant alpha chain with very low diversity \cite{Greenaway2013}. Specific V-J combinations are chosen (TRAV10/TRAJ18 for NKTs and TRAV1-2/TRAJ33 for MAIT) and no nucleotides are inserted in the recombination process of these clonotypes. To see whether these clonotypes could potentially confound our analysis, we searched for published NKT and MAIT sequences in our datasets. $25$ out of the $27$ known  MAIT sequences were found in the datasets at least once ($21$ out of them in the all six individuals), and $8$ out of the $13$ known NKT sequences ($2$ of them in the all six individuals). MAIT and NKT sequences are present in our data, but only a few shared sequences could be explained by them. The majority of shared zero insertion sequences could thus not be attributed to known MAIT or NKT subsets.  

\begin{table*}[p]
\scriptsize
\begin{center}
\begin{tabular}{ | l | l | l }
\hline
	SMART-Mk cap-switching oligonucleotides & \  \\ \hline
	MK-108 & CAGUGGUAUCAACGCAGAGUACNNNNNNUAATGCUNNNNNNUCTT(rG)(rG)(rG)(rG) \\ \hline
	MK-248 & CAGUGGUAUCAACGCAGAGUACNNNNUNNTGGCANNUNNNNNNUCTT(rG)(rG)(rG)(rG) \\ \hline
	MK-253 & CAGUGGUAUCAACGCAGAGUACNNNNUNNTTATGNNUNNNNNNUCTT(rG)(rG)(rG)(rG) \\ \hline
	MK-103 & CAGUGGUAUCAACGCAGAGUACNNNNNNUAACGGUNNNNNNUCTT(rG)(rG)(rG)(rG) \\ \hline
	MK-257 & CAGUGGUAUCAACGCAGAGUACNNNNUNNTTGCGNNUNNNNNNUCTT(rG)(rG)(rG)(rG) \\ \hline
	MK-143 & CAGUGGUAUCAACGCAGAGUACNNNNNNUCAGATUNNNNNNUCTT(rG)(rG)(rG)(rG) \\ \hline
	MK-135 & CAGUGGUAUCAACGCAGAGUACNNNNNNUATGCAUNNNNNNUCTT(rG)(rG)(rG)(rG) \\ \hline
	MK-227 & CAGUGGUAUCAACGCAGAGUACNNNNUNNTAACCNNUNNNNNNUCTT(rG)(rG)(rG)(rG) \\ \hline
	 & \  \\ \hline
	cDNA synthesis primers & \  \\ \hline
	BC\_uni\_R4vvshort & TGGAGTCATTGA    \\ \hline
	TRAC\_R2 & ACACATCAGAATCCTTACTTTG    \\ \hline
	 & \  \\ \hline
	PCR I step primers &  \\ \hline
	Sm1msq & GAGATCTACACGAGTCAGCAGTGGTATCAACGCAG  \\ \hline
	RPbcj1 & CGACTCAGATTGGTACACCTTGTTCAGGTCCTC  \\ \hline
	RPbcj2 & CGACTCAGATTGGTACACGTTTTTCAGGTCCTC \\ \hline
	RPacj & CGACTCAAGTGTGTGGGTCAGGGTTCTGGATAT \\ \hline
	\  & \  \\ \hline
	PCR II step primers & XXXXXX stands for the Truseq index\  \\ \hline
	Sm-out-msq & AATGATACGGCGACCACCGAGATCTACACGAGTCA \\ \hline
	Il-bcj-indX & CAAGCAGAAGACGGCATACGAGATXXXXXXCGACTCAGATTGGTAC \\ \hline
	Il-acj-indX & CAAGCAGAAGACGGCATACGAGATXXXXXXCGACTCAAGTGTGTGG \\ \hline
	 &  \\ \hline
	Custom sequencing primers & \  \\ \hline
	IL-AIRP & ATATCCAGAACCCTGACCCACACACTTGAGTCG \\ \hline
	IL-IRP-b1 & GAGGACCTGAAAAACGTGTACCAATCTGAGTCG \\ \hline
	IL-IRP-b2 & GAGGACCTGAACAAGGTGTACCAATCTGAGTCG \\ \hline
	IL-RP1-msq & ACACGAGTCAGCAGTGGTATCAACGCAGAGTAC \\ \hline
	IL-RP2-b1 & CGACTCAGATTGGTACACGTTTTTCAGGTCCTC \\ \hline
	IL-RP2-b2 & CGACTCAGATTGGTACACCTTGTTCAGGTCCTC \\ \hline
	IL-ARP2 & CGACTCAAGTGTGTGGGTCAGGGTTCTGGATAT \\ \hline
\end{tabular}
\caption{List of primers used}
\end{center}
\end{table*}

\begin{table*}[p]
\footnotesize
\begin{center}
\begin{tabular}{ | l | l | l |  }
\hline
	Alpha chain & \  & \   \\ \hline
	Sample\_id & Number of reads & Number of UMI \\ \hline
	P1\_CD4 & 6566952 & 430915  \\ \hline
	P1\_CD8 & 4620425 & 378044  \\ \hline
	P1\_unpart & 9571058 & 574439  \\ \hline
	P1\_45RO & 4099026 & 431529  \\ \hline
	P2\_CD4 & 4269624 & 941176  \\ \hline
	P2\_CD8 & 4040615 & 561437  \\ \hline
	P2\_unpart & 8213565 & 873546  \\ \hline
	P2\_45RO & 4608991 & 653326  \\ \hline
	Q1\_CD4 & 3894188 & 653649  \\ \hline
	Q1\_CD8 & 3201067 & 589757  \\ \hline
	Q1\_unpart & 8360990 & 1091786  \\ \hline
	Q1\_45RO & 3587344 & 687916  \\ \hline
	Q2\_CD4 & 3877893 & 828573  \\ \hline
	Q2\_CD8 & 3880048 & 825539  \\ \hline
	Q2\_unpart & 9159719 & 1215155  \\ \hline
	Q2\_45RO & 3890664 & 834828  \\ \hline
	S1\_CD4 & 4655514 & 734158  \\ \hline
	S1\_CD8 & 1009038 & 219433  \\ \hline
	S1\_unpart & 3191701 & 621723  \\ \hline
	S1\_45RO & 4977466 & 495057  \\ \hline
	S2\_CD4 & 11727155 & 761495  \\ \hline
	S2\_CD8 & 12436797 & 468345  \\ \hline
	S2\_unpart & 11135704 & 610105  \\ \hline
	S2\_45RO & 9064981 & 633362  \\ \hline
	\  & \  & \  \\ \hline
	Beta chain & \  & \   \\ \hline
	Sample\_id & Number of reads & Number of UMI \\ \hline
	P1\_CD4 & 3757755 & 759270  \\ \hline
	P1\_CD8 & 3565384 & 517737  \\ \hline
	P1\_unpart & 7429601 & 955106  \\ \hline
	P1\_45RO & 4036708 & 695379  \\ \hline
	P2\_CD4 & 3042278 & 449048  \\ \hline
	P2\_CD8 & 3438238 & 477696  \\ \hline
	P2\_unpart & 8144134 & 817306  \\ \hline
	P2\_45RO & 4598733 & 578663  \\ \hline
	Q1\_CD4 & 3694288 & 673037  \\ \hline
	Q1\_CD8 & 4586088 & 758201  \\ \hline
	Q1\_unpart & 6511237 & 1060251  \\ \hline
	Q1\_45RO & 3171012 & 664732  \\ \hline
	Q2\_CD4 & 3066472 & 605062  \\ \hline
	Q2\_CD8 & 3389029 & 691438  \\ \hline
	Q2\_unpart & 7256515 & 1241753  \\ \hline
	Q2\_45RO & 3110044 & 667997  \\ \hline
	S1\_CD4 & 3510759 & 722883  \\ \hline
	S1\_CD8 & 3162597 & 489393  \\ \hline
	S1\_unpart & 7019324 & 1181194  \\ \hline
	S1\_45RO & 3363725 & 574876  \\ \hline
	S2\_CD4 & 4034384 & 717023  \\ \hline
	S2\_CD8 & 4267632 & 546529  \\ \hline
	S2\_unpart & 7093628 & 875357  \\ \hline
	S2\_45RO & 2848644 & 526765  \\ \hline
	Memory\_aged19 & 7486248 & 424156  \\ \hline
	Naïve\_aged19 & 9166800 & 932396  \\ \hline
	Memory\_aged51 & 4376542 & 366646  \\ \hline
	Naïve\_aged51 & 4115592 & 602950  \\ \hline
	Memory\_aged57 & 5743372 & 476395  \\ \hline
	Naïve\_aged57 & 5227973 & 358245  \\ \hline
	Cord\_blood & 8015355 & 1803557  \\ \hline
\end{tabular}
\caption{Number of reads and UMI in each sample.}
\end{center}
\end{table*}

 \begin{table*}[p]
 \tiny
 \begin{center}
\begin{tabular}{ | l | l | l | l | }
\hline
	Sample id & fraction of 0 ins in top 2000 & Naive,\% & Age, years \\ \hline
	A2-i132 & 0.015056135255 & 73.7 & 6 \\ \hline
	A2-i131 & 0.010037196444 & 43 & 9 \\ \hline
	A2-i136 & 0.027691639038 & 40 & 10 \\ \hline
	A2-i129 & 0.0108412940125 & 57 & 11 \\ \hline
	A2-i134 & 0.021007545075 & 68 & 16 \\ \hline
	A2-i133 & 0.0119257041822 & 60.9 & 16 \\ \hline
	A4-i194 & 0.013765206508 & 55 & 20 \\ \hline
	A4-i195 & 0.0119673129492 & 59 & 21 \\ \hline
	A4-i191 & 0.01637900271 & 45 & 22 \\ \hline
	A4-i192 & 0.012716977224 & 56 & 24 \\ \hline
	A4-i189 & 0.012839842368 & 44 & 25 \\ \hline
	A6-I201ob & 0.0091925381272 & NA & 30 \\ \hline
	A3-i110 & 0.0078554903232 & 36.4 & 34 \\ \hline
	A3-i101 & 0.0107838068688 & 55 & 36 \\ \hline
	A4-i101 & 0.0090257537105 & 27 & 36 \\ \hline
	A4-i102 & 0.00628983345724 & 27.6 & 37 \\ \hline
	A3-i107 & 0.00851643362094 & 43 & 39 \\ \hline
	A4-i107 & 0.0064344051544 & 26 & 39 \\ \hline
	A3-i106 & 0.016159136094 & 39.4 & 43 \\ \hline
	A3-i102 & 0.0107591339774 & 27.3 & 43 \\ \hline
	A4-i110 & 0.018164859228 & 40 & 43 \\ \hline
	A4-i106 & 0.00642081990976 & 31 & 43 \\ \hline
	A5-S23 & 0.0046042762969 & 21.3 & 50 \\ \hline
	A5-S24 & 0.0061143105585 & 29.9 & 50 \\ \hline
	A6-I160 & 0.008621670788 & 38.9 & 51 \\ \hline
	A5-S21 & 0.0086245934928 & 51.3 & 51 \\ \hline
	A6-I215ob & 0.00819076572358 & NA & 51 \\ \hline
	A5-S22 & 0.00695571384444 & 48.5 & 51 \\ \hline
	A6-I150 & 0.0061129801278 & NA & 51 \\ \hline
	A5-S20 & 0.00387005779589 & 25 & 51 \\ \hline
	A5-S19 & 0.0080402564192 & 41.2 & 55 \\ \hline
	A4-i185 & 0.0085319088075 & 29.6 & 61 \\ \hline
	A4-i186 & 0.00532914538306 & 14.6 & 61 \\ \hline
	A4-i184 & 0.00405847825812 & 21 & 61 \\ \hline
	A4-i188 & 0.00663226556694 & 18 & 61 \\ \hline
	A4-i128 & 0.0058717051432 & 23 & 62 \\ \hline
	A4-i125 & 0.00476704046791 & 4.5 & 64 \\ \hline
	A4-i124 & 0.00394006128853 & 16.3 & 66 \\ \hline
	A2-i141 & 0.0060990185169 & 30 & 71 \\ \hline
	A2-i140 & 0.0081195988401 & 47 & 73 \\ \hline
	A2-i138 & 0.00507840452028 & 6.7 & 74 \\ \hline
	A2-i139 & 0.008749966888 & 28.2 & 75 \\ \hline
	A4-i122 & 0.00606575047668 & 33 & 85 \\ \hline
	A3-i145 & 0.004749303571 & 37 & 86 \\ \hline
	A4-i132 & 0.0034771649962 & 14.5 & 87 \\ \hline
	A4-i183 & 0.00723588404502 & 24.6 & 87 \\ \hline
	A3-i150 & 0.0037069726895 & 13.3 & 87 \\ \hline
	A6-I214ob & 0.0046188525124 & 21 & 88 \\ \hline
	A5-S10 & 0.007023235658 & NA & 89 \\ \hline
	A4-i118 & 0.00512286685575 & 54 & 89 \\ \hline
	A4-i127 & 0.005589445878 & 12.7 & 90 \\ \hline
	A5-S9 & 0.00642820638494 & 26.5 & 90 \\ \hline
	A6-I211ob & 0.00432554146357 & 8.4 & 91 \\ \hline
	A5-S8 & 0.00421932231855 & 4.5 & 92 \\ \hline
	A5-S7 & 0.0078096377085 & 4.7 & 92 \\ \hline
	A6-I210ob & 0.00368734455504 & 7.4 & 92 \\ \hline
	A6-I208ob & 0.0045677109953 & 8.7 & 93 \\ \hline
	A5-S4 & 0.0046450251048 & 30.8 & 93 \\ \hline
	A6-I207ob & 0.0044350512973 & 27.6 & 94 \\ \hline
	A6-I206ob & 0.0061812657375 & 6.2 & 95 \\ \hline
	A6-I205ob & 0.00481739413682 & 7.5 & 95 \\ \hline
	A5-S3 & 0.0040549739527 & 12.4 & 98 \\ \hline
	A6-I204ob & 0.00431740407138 & 10.3 & 99 \\ \hline
	A5-S2 & 0.00486991171424 & 15.5 & 100 \\ \hline
	A5-S1 & 0.00541415235339 & NA & 103 \\ \hline
\end{tabular}
 \caption{Ageing data used for Fig. 4 and exponential decay fits. Percentage of the naive T-cells defined using flow cytometry, see \cite{Britanova2014} for details. }
 \end{center}
 \end{table*}

\end{document}